\documentclass[review]{elsarticle}

\usepackage{hyperref}
\usepackage{amsmath}
\usepackage{color}

\journal{Ecological Complexity}

\begin{document}

\begin{frontmatter}

\title{Evolutionary dynamics and competition stabilize three-species predator-prey 
       communities}
%\title{The interplay between evolutionary dynamics and competition stabilizes 
%biodiversity in a three-species predator-prey community}
%\tnotetext[mytitlenote]{Fully documented templates are available in the 
%elsarticle package on \href{http://www.ctan.org/tex-archive/macros/latex/contrib/elsarticle}{CTAN}.}

%\author{Sheng Chen, \fnref{myfootnote}}
%\fntext[myfootnote]{csheng@vt.edu}
%\author{Uli, \fnref{ulifootnote}}
%\address{Radarweg 28, Amsterdam}
%\fntext[myfootnote]{uli@vt.edu}

\author[mymainaddress]{Sheng Chen}
\ead{csheng@vt.edu}

\author[mysecondaryaddress]{Ulrich Dobramysl}
\ead{u.dobramysl@gurdon.cam.uc.uk}
\cortext[mycorrespondingauthor]{Corresponding author: Ulrich Dobramysl, \href{mailto:u.dobramysl@gurdon.cam.ac.uk}{u.dobramysl@gurdon.cam.ac.uk}, Wellcome Trust / Cancer Research UK Gurdon Institute, 
         University of Cambridge, Cambridge CB2 1QN, United Kingdom}

\author[mymainaddress]{Uwe C. T\"auber}
\ead{tauber@vt.edu}

\address[mymainaddress]{Department of Physics (MC 0435) and Center for Soft Matter 
         and Biological Physics, Robeson Hall, 850 West Campus Drive, Virginia Tech, 
         Blacksburg, Virginia 24061, USA}
\address[mysecondaryaddress]{Wellcome Trust / Cancer Research UK Gurdon Institute, 
         University of Cambridge, Cambridge CB2 1QN, United Kingdom}

\begin{abstract}
We perform individual-based Monte Carlo simulations in a community consisting of 
two predator species competing for a single prey species, with the purpose of 
studying biodiversity stabilization in this simple model system.
Predators are characterized with predation efficiency and death rates, to which 
Darwinian evolutionary adaptation is introduced. 
Competition for limited prey abundance drives the populations' optimization with
respect to predation efficiency and death rates.
We study the influence of various ecological elements on the final state, finding
that both indirect competition and evolutionary adaptation are insufficient to 
yield a stable ecosystem. 
However, stable three-species coexistence is observed when direct interaction 
between the two predator species is implemented.
\end{abstract}

\begin{keyword}
%\texttt{elsarticle.cls}\sep \LaTeX\sep Elsevier \sep template
%\MSC[2010] 00-01\sep  99-00
Darwinian evolution \sep interspecific competition \sep Lotka-Volterra model \sep 
multi-species coexistence \sep character displacement 
\end{keyword}
\end{frontmatter}

\section{Introduction}
\label{sec1}

%\paragraph{Installation} If the document class \emph{elsarticle} is not 
%available on your computer, you can download and install the system package 
%\emph{texlive-publishers} (Linux) or install the \LaTeX\ package \emph{elsarticle} 
%using the package manager of your \TeX\ installation, which is typically \TeX\ Live or Mik\TeX.
Ever since Darwin first introduced his theory that interspecific competition 
positively contributes to ecological character displacement and adaptive divergence 
\cite{d1859}, debates abounded about its importance in biodiversity. 
Character displacement is considered to occur when a phenotypical feature of the 
animal \cite{bwe1956}, which could be morphological, ecological, behavioral, or 
physiological, beak size for example, is shifted in a statistically significant 
manner due to the introduction of a competitor \cite{dj1992, mt1992}. 
One example of ecological character displacement is that the body size of an island 
lizard species becomes reduced on average upon the arrival of a second, competing 
lizard kind \cite{j2002}.
Early observational and experimental studies of wild animals provided support for 
Darwinian evolutionary theory \cite{l1947, bwe1956}. 
One famous observation related to finches, whose beak size would change in 
generations because of competition \cite{l1947}. 
However, recent studies using modern genetic analysis techniques do not find genetic 
changes to the same extent as the phenotypic break change, thereby casting doubt on 
Darwin's observational studies \cite{g1975, aw1982}. 
Another concern with experiments on birds or other animal species is that they may 
live for decades, rendering this sort of study too time-consuming. 
Evolutionary theory is based on the assumption that interspecific competition occurs 
mostly between closely related species because they share similar food resources, 
thus characters exploiting new resources are preferred. 
Ecologists perform experiments with wild animals by introducing a second competing 
species and recording their observable characters including the body size, beak 
length, and others \cite{aw1982, j2002}. 
Unfortunately, direct control over natural ecosystems is usually quite limited; for 
example, ecological character displacement with wild animals cannot be shut down at 
will in natural habitats. 
However, this is easily doable in carefully designed computer simulations. 

Game theory has a long history in the study of biological problems \cite{j1982}. 
Among all the mathematical models of studying biodiversity in ecology, the 
Lotka--Volterra (LV) \cite{a1920, v1926} predator-prey model may rank as possibly 
the simplest one. 
Only one predator and one prey species are assumed to exist in the system. 
Individuals from each species are regarded as simple particles with their reaction 
rates set uniformly and spatially homogeneous. 
They display three kinds of behaviors which are influenced by pre-determined 
reaction rates: prey particles may reproduce, predator particles can spontaneously 
die, and predators may remove a prey particle and simultaneously reproduce. 
This simple LV model kinetics may straightforwardly be implemented on a regular 
lattice (usually square in two or cubic in three dimensions) to simulate situations 
in nature, where stochasticity as well as spatio-temporal correlations play an 
important role \cite{hnak1992}--\cite{su2016}. 
It is observed in such spatial stochastic LV model systems that predator and prey 
species may coexist in a quasi-stable steady state where both populations reach 
non-zero densities that remain constant in time; here, the population density is 
defined as the particle number of one species divided by the total number of lattice 
sites. 
Considering that the original LV model contains only two species, we here aim to 
modify it to study a multi-species system. 
We note that there are other, distinct well-studied three-species models, including 
the rock-paper-scissors model \cite{rps,ummu2017}, which is designed to study cyclic 
competitions, and a food-chain-like three-species model \cite{hn2014}, as well as 
more general networks of competing species \cite{ummu2017}, all of which contain 
species that operate both as a predator and a prey. 
In this paper we mainly focus on predator-prey competitions, where any given species
plays only one of those ecological roles.

Compared with the original LV model, we introduce one more predator into the system 
so that there are two predator species competing for the same prey. 
We find that even in a spatially extended and stochastic setting, the `weaker' of 
the two predator species will die out fast if all reaction rates are fixed.
Afterwards the remaining two species form a standard LV system and approach stable 
steady-state densities. 
Next we further modify the model by introducing evolutionary adaptation 
\cite{uu2013}. 
We also add a positive lower bound to the predator death rates in order to avoid 
`immortal' particles. 
Finally, we incorporate additional direct competition between predator individuals.
Stable multiple-species coexistence states are then observed in certain parameter 
regions, demonstrating that adaptive `evolution' in combination with direct 
competition between the predator species facilitate ecosystem stability. 
Our work thus yields insight into the interplay between evolutionary processes and 
inter-species competition and their respective roles to maintain biodiversity.

\section{Stochastic lattice Lotka--Volterra Model with fixed reaction rates}
\label{sec2}

\subsection{Model description}
\label{subsec2.1}

We spatially extend the LV model by implementing it on a two-dimensional square 
lattice with linear system size $L=512$. 
It is assumed that there are three species in the system: two predator species $A$, 
$B$, and a single prey species $C$. 
Our model ignores the detailed features and characters of real organisms, and 
instead uses simple `particles' to represent the individuals of each species. 
These particles are all located on lattice sites in a two-dimensional space with 
periodic boundary conditions (i.e., on a torus) to minimize boundary effects. 
Site exclusion is imposed to simulate the natural situation that the local 
population carrying capacity is finite: 
Each lattice site can hold at most one particle, i.e., is either occupied by one 
`predator' $A$ or $B$, occupied by one `prey' $C$, or remains empty. 
This simple model partly captures the population dynamics of a real ecological 
system because the particles can predate, reproduce, and spontaneously die out; 
these processes represent the three main reactions directly affecting population 
number changes.
There is no specific hopping process during the simulation so that a particle will 
never spontaneously migrate to other sites. 
However, effective diffusion is brought in by locating the offspring particles on 
the neighbor sites of the parent particles in the reproduction process 
\cite{miu2006, su2016}. 
The stochastic reactions between neighboring particles are described as follows: 
\begin{equation}
\begin{split}
\label{lvreac21}
	A \xrightarrow{\mu_A} \emptyset \, &, \quad  
	B \xrightarrow{\mu_B} \emptyset \, , \\
	A + C \xrightarrow{\lambda_A } A + A \, &, \quad 
	B + C \xrightarrow{\lambda_B } B + B \, , \quad \\
	C & \xrightarrow{\sigma} C+C \, .
\end{split}
\end{equation}
The `predator' A (or B) may spontaneously die with decay rate 
$\mu_A\, (\mu_B) > 0$. 
The predators may consume a neighboring prey particle $C$, and simultaneously 
reproduce with `predation' rate  $\lambda_{A/B}$, which is to replace $C$ with a new 
predator particle in the simulation. 
In nature, predation and predator offspring production are separate processes. 
But such an explicit separation would not introduce qualitative differences in a 
stochastic spatially extended system in dimensions $d<4$ \cite{miu200602}. 
When a prey particle has an empty neighboring site, it can generate a new offspring
prey individual there with birth rate $\sigma > 0$. 
Note that a separate prey death process $C\rightarrow 0$ can be trivially described 
by lowering the prey reproduction rate and is therefore not included. 
We assume asexual reproduction for all three species, i.e., only one parent particle 
is involved in the reproduction process. 
Each species consists of homogeneous particles with identical reaction rates. 
Predator species $A$ and $B$ may be considered as close relatives since they display 
similar behavior (decay, predation and reproduction, effective diffusion) and most 
importantly share the same mobile food source $C$. 
For now, we do not include evolution in the reproduction processes, therefore all offspring particles are exact clones of their parents. 
We are now going to show that these two related predator species can never coexist.

\subsection{Mean-field rate equations}
\label{subsec2.2}

The mean-field approximation ignores spatial and temporal correlations and 
fluctuations, and instead assumes the system to be spatially well-mixed. 
We define $a(t)$ and $b(t)$ as the predators' population densities and $c(t)$ as the 
prey density. 
Each predator population decreases exponentially with death rate $\mu$, but
increases with the predation rate $\lambda$ and prey density $c(t)$.
The prey population $c(t)$ increases exponentially with its reproduction rate 
$\sigma$, but decreases as a function of the predator population densities.  
The mean-field rate equations consequently read
\begin{equation}
\begin{split}
	\frac{d a(t)}{dt} &= -\mu_A a(t) + \lambda_A a(t)c(t)\,, \\
	\frac{d b(t)}{dt} &= -\mu_B b(t) + \lambda_B b(t)c(t)\,, \\
	\frac{d c(t)}{dt} &= \sigma c(t)\bigg[1-\frac{a(t)+b(t)+c(t)}{K}\bigg] 
        - \lambda_A a(t)c(t) -\lambda_B b(t)c(t)\,.
\end{split}
\label{lvreac22}
\end{equation}
$K > 0$ represents the finite prey carrying capacity. 
In order to obtain stationary densities, the left-side derivative terms are set to 
zero. 
The ensuing (trivial) extinction fixed points are: (1) $a = b = c = 0$; 
(2) $a = b = 0$, $c = K$; (3) for $\mu_A<\lambda_AK$: 
$a=\frac{\sigma(\lambda_AK-\mu_A)}{\lambda_A(\lambda_AK+\sigma)}$, $b=0$, 
$c=\mu_A/\lambda_A$; (4) for $\mu_B<\lambda_BK$: $a=0$, 
$b=\frac{\sigma(\lambda_BK-\mu_B)}{\lambda_B(\lambda_BK+\sigma)}$, 
$c=\mu_B/\lambda_B$.
When $\mu_A/\lambda_A \neq \mu_B/\lambda_B$, there exists no three-species 
coexistence state. 
Yet in the special situation $\mu_A/\lambda_A = \mu_B/\lambda_B$, another line of
fixed points emerges: $(\frac{\sigma}{K}+\lambda_A)a+(\frac{\sigma}{K}+\lambda_B)b
+\frac{\sigma}{K}c=\sigma$, $c=\mu_A/\lambda_A =\mu_B/\lambda_B$. 

\subsection{Lattice Monte Carlo simulation results}
\label{subsec2.3}

\begin{figure*}[]
\centering
\includegraphics[width=0.32\columnwidth]{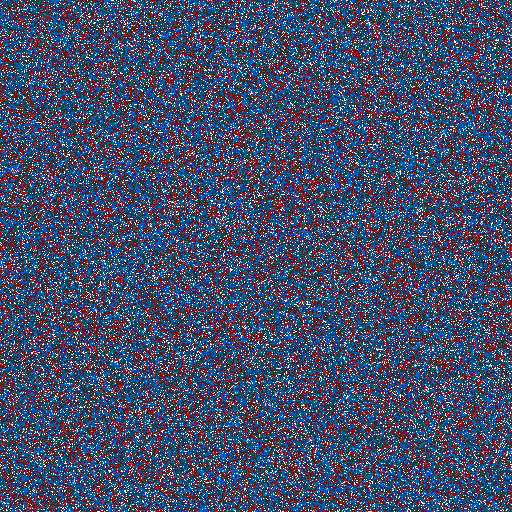}
\includegraphics[width=0.32\columnwidth]{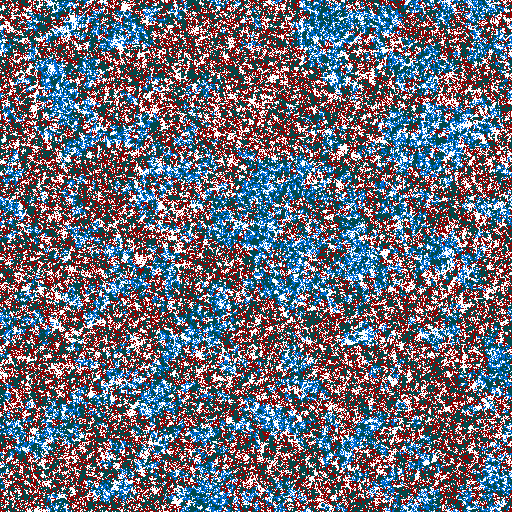}
\includegraphics[width=0.32\columnwidth]{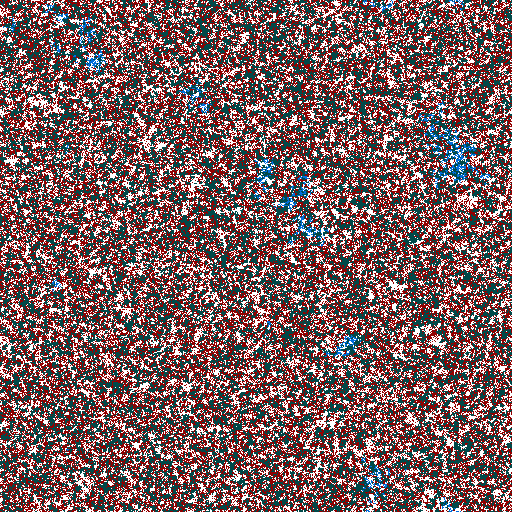}
\caption{Snapshots of the spatial particle distribution for a single Monte Carlo 
	simulation run of a stochastic predator-predator-prey Lotka--Volterra model 
	on a $512\times 512$ square lattice with periodic boundary conditions at 
	(from left to right) $t = 0$ Monte Carlo Steps (MCS), $t = 10\,000$ MCS, and 
	$t = 50\,000$ MCS, with predation rates $\lambda_A=0.5$, $\lambda_B=0.5$, 
	predator death rates $\mu_A=0.126$, $\mu_B=0.125$, and prey reproduction 
	rate $\sigma=1.0$. Only at most one particle per lattice site is allowed.
	Predator particles $A$ are indicated in blue, predators $B$ in red, and 
	prey $C$ in dark green, while empty sites are shown in white.}
\label{fig1}
\end{figure*}

\begin{figure*}[t]
\centering
\includegraphics[width=0.9\columnwidth]{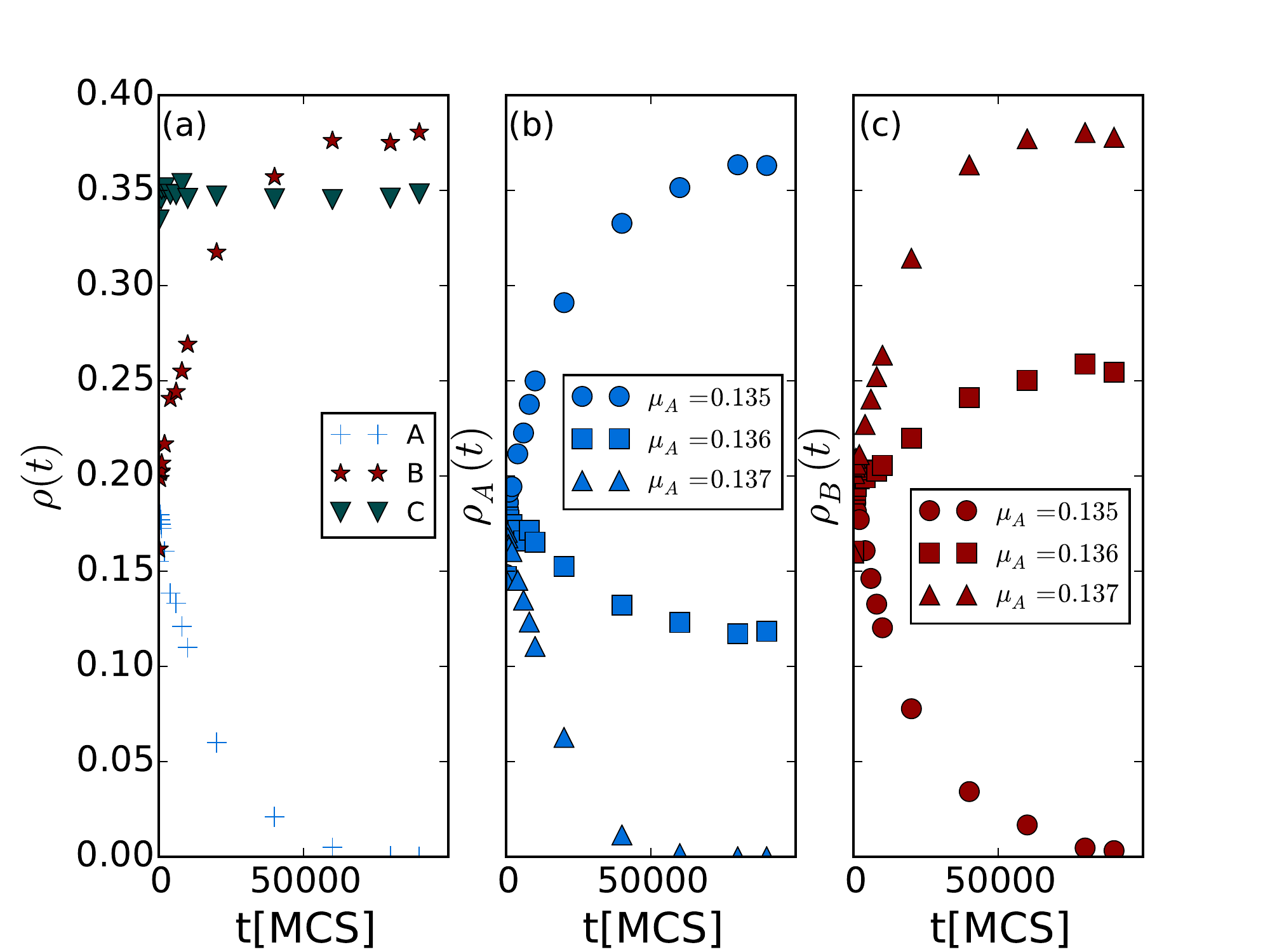}  
\caption{The two predator species cannot coexist in Monte Carlo simulations of the 
	two-predator-one-prey model with fixed reaction rates. 
	(a) Time evolution of the population densities with fixed reaction rates: 
 	predation rates $\lambda_A=0.5$, $\lambda_B=0.5$, predator death rates 
	$\mu_A=0.126$, $\mu_B=0.125$, and prey reproduction rate $\sigma=1.0$; 
	(b,c) temporal evolution of the population densities $\rho_A(t)$ and 
	$\rho_B(t)$ with fixed $\lambda_A=0.55$, $\lambda_B=0.5$, $\mu_B=0.125$, and 
	$\mu_A$ varying from $0.135$, $0.136$, to $0.137$. 
	The curves in (b) and (c) sharing the same markers are from the same (single)
	simulation runs.}
\label{fig2}
\end{figure*}

In the stochastic lattice simulations, population densities are defined as the 
particle numbers for each species divided by the total number of lattice sites 
($512\times512$).
We prepare the system so that the starting population densities of all three species 
are the same, here set to $0.3$ (particles/lattice site), and the particles are 
initially randomly distributed on the lattice.  
The system begins to leave this initial state as soon as the reactions start and the 
ultimate stationary state is only determined by the reaction rates, independent of 
the system's initialization. 
We can test the simulation program by setting the parameters as 
$\lambda_A = \lambda_B = 0.5$ and $\mu_A = \mu_B = 0.125$. 
Since species $A$ and $B$ are now exactly the same, they coexist with an equal 
population density in the final stable state, as indeed observed in the simulations. 
We increase the value of $\mu_A$ by $0.001$ so that predator species $A$ is more 
likely to die than $B$. 
Fig.~\ref{fig1} shows the spatial distribution of the particles at $0$, $10\,000$, 
and $50\,000$ Monte Carlo Steps (MCS, from left to right), indicating sites occupied 
by $A$ particles in blue, $B$ in red, $C$ in green, and empty sites in white. 
As a consequence of the reaction scheme (\ref{lvreac21}), specifically the clonal
offspring production, surviving particles in effect remain close to other 
individuals of the same species and thus form clusters.
After initiating the simulation runs, one typically observes these clusters to 
emerge quite quickly; as shown in Fig.~\ref{fig1}, due to the tiny difference 
between the death rates $\mu_A - \mu_B > 0$, the `weaker' predator species $A$ 
gradually decreases its population number and ultimately goes extinct.
Similar behavior is commonly observed also with other sets of parameters: 
For populations with equal predation rates, only the predator species endowed 
with a lower spontaneous death rate will survive. 
Fig.~\ref{fig2}(a) records the temporal evolution of the three species' population 
densities. 
After about $60\,000$ MCS, predator species $A$ has reached extinction, while the 
other two populations eventually approach non-zero constant densities. 
With larger values of $\mu_A$ such as $0.127$ or $0.13$, species $A$ dies out within 
a shorter time interval; the extinction time increases with diminishing death rate
difference $|\mu_A-\mu_B|$.  

In Figs.~\ref{fig2}(b) and (c), we set $\lambda_A=0.55$, $\lambda_B=0.5$, 
$\mu_B=0.125$, and various values of $\mu_A > 0.13$. 
The larger rate $\lambda_A$ gives species $A$ an advantage over $B$ in the predation 
process, while the bigger rate $\mu_A$ enhances the likelihood of death for $A$ as
compared to $B$. 
Upon increasing $\mu_A$ from $0.135$ to $0.137$, we observe a phase transition from 
species $B$ dying out to $A$ going extinct in this situation with competing 
predation and survival advantages.
When $\mu_A$ thus exceeds a certain critical value (in this example near $0.136$), 
the disadvantages of high death rates cannot balance the gains due to a
more favorable predation efficiency; hence predator species $A$ goes extinct.
In general, whenever the reaction rates for predator species $A$ and $B$ are not 
exactly the same, either $A$ or $B$ will ultimately die out, while the other species
remains in the system, coexisting with the prey $C$. 
This corresponds to actual biological systems where two kinds of animals share 
terrain and compete for the same food. 
Since there is no character displacement occurring between generations, the weaker 
species' population will gradually decrease. 
This trend cannot be turned around unless the organisms improve their capabilities 
or acquire new skills to gain access to other food sources; either change tends to 
be accompanied by character displacements 
\cite{gg2006, aald2009, ytprlj2014, jmxl2016}.

\begin{figure*}[t]
\centering
\includegraphics[width=0.9\columnwidth]{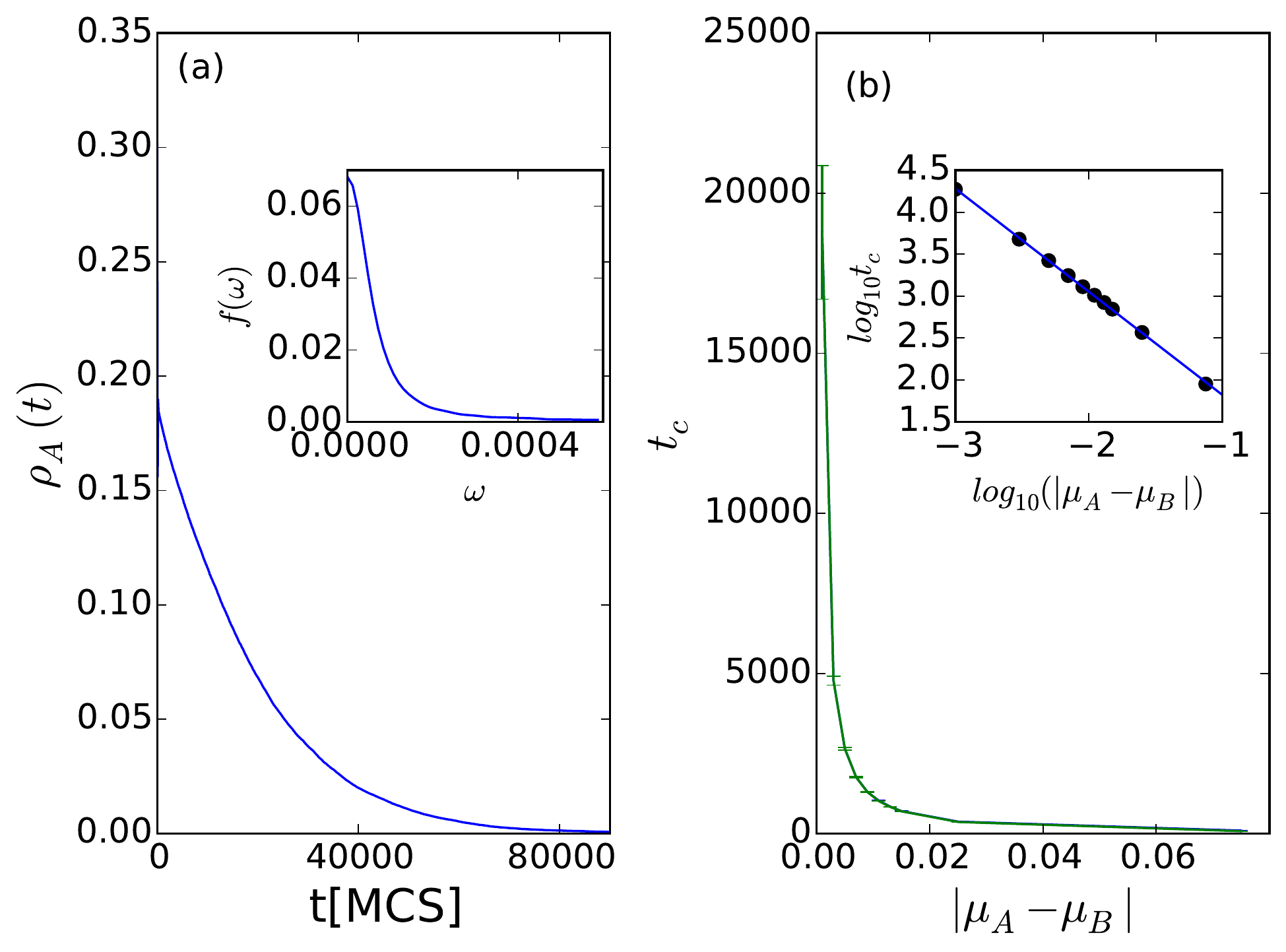}  
\caption{Characteristic decay time of the weaker predator species measured in Monte 
	Carlo simulations of the two-predator-one-prey model with fixed reaction 
	rates. 
	(a) Main panel: temporal evolution of the predator population density 
	$\rho_A(t)$ with predation rates $\lambda_A=0.5$, $\lambda_B=0.5$, predator 
	death rates $\mu_A=0.126$, $\mu_B=0.125$, and prey reproduction rate 
	$\sigma=1.0$. Inset: Fourier transform amplitude $f(\omega)$ of the predator
	density time series $\rho_A(t)$.
	(b) Main panel: characteristic decay time $t_c$ as obtained from the peak 
	width of $f(\omega)$, versus the death rate difference $|\mu_A-\mu_B|$, with 
	all other reaction rates fixed as in (a). Inset: the black dots show the data
	points $\log_{10}t_c$ versus $\log_{10}(|\mu_A-\mu_B|)$, while the blue 
	straight line with slope $-1.23\pm 0.01$ is inferred from linear regression.}
\label{fig3}
\end{figure*}

In order to quantitatively investigate the characteristic time for the weaker 
predator species to vanish, we now analyze the relation between the relaxation time 
$t_c$ of the weaker predator species ($A$ here) and the difference of death rates 
$|\mu_A-\mu_B|$ under the condition that $\lambda_A=\lambda_B$.
Fig.~\ref{fig2}(a) indicates that prey density (green triangles) reaches its 
stationary value much faster than the predator populations. 
When $|\mu_A-\mu_B|$ becomes close to zero, the system returns to a two-species 
model, wherein the relaxation time of the prey species $C$ is finite. 
However, the relaxation time of either predator species would diverge because it 
takes longer for the stronger species to remove the weaker one when they become very
similar in their death probabilities. 
Upon rewriting eqs.~\eqref{lvreac22} for $\lambda_A=\lambda_B$ by replacing the prey 
density $c(t)$ with its stationary value $\mu_B/\lambda_B$, we obtain a linearized 
equation for the weaker predator density: $\frac{da(t)}{dt}=-|\mu_A-\mu_B| a(t)$, 
describing exponential relaxation with decay time $t_c=1/|\mu_A-\mu_B|$.

We further explore the relation between the decay rate of the weak species 
population density and the reaction rates through Monte Carlo simulations. 
Fig.~\ref{fig3}(a) shows an example of the weaker predator $A$ population density 
decay for fixed reaction rates $\lambda_A=0.5$, $\lambda_B=0.5$, $\mu_A=0.126$, 
$\mu_B=0.125$, and $\sigma=1.0$, and in the inset also the corresponding Fourier 
amplitude $f(\omega) = | \int e^{- i \omega t} \, \rho_A(t) \, dt |$ that is 
calculated by means of the fast Fourier transform algorithm. 
Assuming an exponential decay of the population density according to 
$\rho_A(t)\sim e^{-t/t_c}$, we identify the peak half-width at half maximum with the 
inverse relaxation time $1/t_c$. 
For other values of $\mu_A>0.125$, the measured relaxation times $t_c$ for the 
predator species $A$ are plotted in Fig.~\ref{fig3}(b). 
We also ran simulations for various parameter values $\mu_A<0.125$, for which the 
predator population $B$ would decrease toward extinction instead of $A$, and 
measured the corresponding relaxation time for $\rho_B(t)$, plotted in 
Fig.~\ref{fig3}(b) as well. 
The two curves overlap in the main panel of Fig.~\ref{fig3}(b), confirming that 
$t_c$ is indeed a function of $|\mu_A-\mu_B|$ only. 
The inset of Fig.~\ref{fig3}(b) demonstrates a power law relationship  
$t_c \sim |\mu_A-\mu_B|^{- z \nu}$ between the relaxation time and the reaction rate
difference, with exponent $z \nu \approx -1.23\pm 0.01$ as inferred from the slope 
in the double-logarithmic graph via simple linear regression.
This value is to be compared with the corresponding exponent 
$z \nu \approx 1.295\pm 0.006$ for the directed percolation (DP) universality class
\cite{dp1996}. 
Directed percolation \cite{dp} represents a class of models that share identical 
values of their critical exponents at their phase transition points, and is expected
to generically govern the critical properties at non-equilibrium phase transitions 
that separate active from inactive, absorbing states \cite{noneq1, uct2014}. 
Our result indicates that the critical properties of the two-predator-one-prey model 
with fixed reaction rates at the extinction threshold of one predator species 
appear to also be described by the DP universality class.  

\begin{figure*}[]
\centering
\includegraphics[width=1.0\columnwidth]{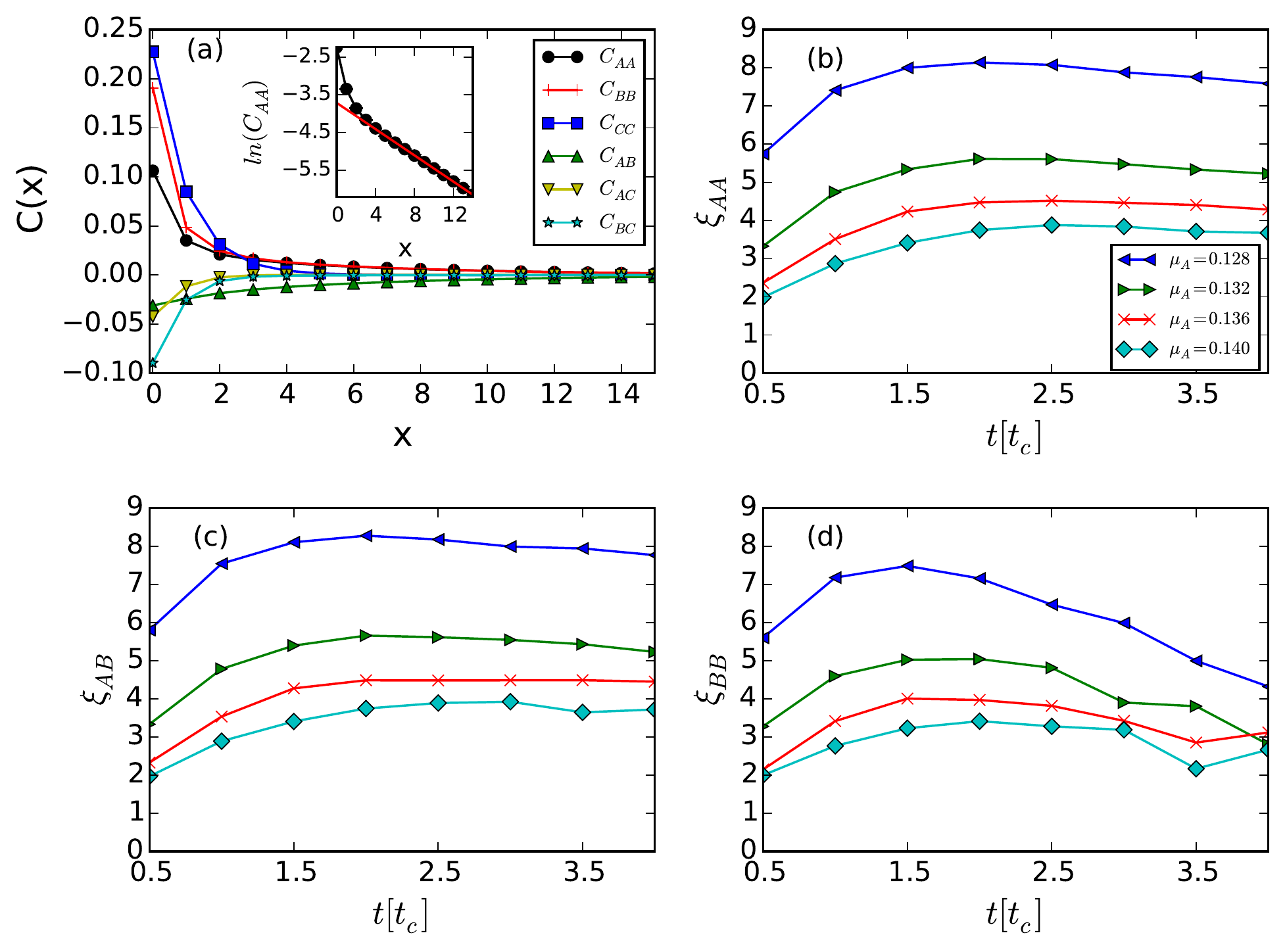}  
\caption{Time evolution for correlation lengths during Monte Carlo simulations of 
	the two-predator-one-prey model with fixed reaction rates. 
	(a) Main panel: correlation functions $C(x)$ after the system has evolved for
	one half of the relaxation time $0.5 t_c \approx 2386$ MCS, with reaction 
	rates $\lambda_A=0.5$, $\lambda_B=0.5$, $\mu_A=0.128$, $\mu_B=0.125$, and 
	$\sigma=1.0$. Inset: $\ln(C_{AA})$ with a simple linear regression of the 
	data points with $x\in [4, 14]$ (red straight line) that yields the 
	characteristic correlation decay length $\xi_{AA}\approx 5.8$. 
	(b,c,d) Measured correlation lengths $\xi_{AA}$, $\xi_{AB}$, and $\xi_{BB}$ 
	as function of the system evolution time t relative to $t_c$, with reaction 
	rates as in (a) except (top to bottom) $\mu_A=0.128$ (blue left triangles), 
	$0.132$ (green right triangles), $0.136$ (red crosses), and $0.140$ (cyan 
	diamonds).}
\label{fig4}
\end{figure*}

As already shown in Fig.~\ref{fig1}, individuals from each species form clusters in 
the process of the stochastically occurring reactions (\ref{lvreac22}). 
The correlation lengths $\xi$, obtained from equal-time correlation functions $C(x)$,
characterize the average sizes of these clusters. 
The definition of the correlation functions between the different species 
$\alpha, \beta = A, B, C$ is 
$C_{\alpha\beta}(x)=\langle n_\alpha (x) n_\beta (0) \rangle - 
 \langle n_\alpha (x) \rangle \langle n_\beta (0) \rangle$, where $n_\alpha (x)=0,1$ 
denotes the local occupation number of species $\alpha$ at site $x$. 
First choosing a lattice site, and then a second site at distance $x$ away, we note
that the product $n_\alpha(x)n_\beta(0)=1$ only if a particle of species $\beta$ is 
located on the first site, and a particle of species $\alpha$ on the second site; 
otherwise the product equals $0$. 
We then average over all sites to obtain $\langle n_\alpha (x) n_\beta (0)\rangle$. 
$\langle n_\alpha (x)\rangle$ represents the average population density of species 
$\alpha$. 

In our Monte Carlo simulations we find that although the system has not yet reached 
stationarity at $0.5\,t_c$, its correlation functions do not vary appreciably during
the subsequent time evolution.
This is demonstrated in Figs.~\ref{fig4}(b-d) which show the measured correlation 
lengths from $0.5\,t_c$ to $3.75\,t_c$, during which time interval the system 
approaches its quasi-stationary state. 
The main panel in Fig.~\ref{fig4}(a) shows the measured correlation functions after 
the system has evolved for $0.5\,t_c \approx 2386$ MCS, with predator $A$ death
rate $\mu_A=0.128$. 
Individuals from the same species are evidently spatially correlated, as indicated 
by the positive values of $C_{\alpha\alpha}$. 
Particles from different species, on the other hand, display anti-correlations. 
The inset demonstrates exponential decay: $C_{AA}(x)\sim e^{-|x|/\xi_{AA}}$, where 
$\xi_{AA}$ is obtained from linear regression of $ln(C_{AA}(x))$. 
In the same manner, we calculate the correlation length $\xi_{AA}$, $\xi_{BB}$, and 
$\xi_{AB}$ for every $0.5\,t_c$ the system evolves, for different species $A$ death 
rates $\mu_A=0.128$, $0.132$, $0.136$, and $0.140$, respectively. 
Fig.~\ref{fig4}(b) shows that predator $A$ clusters increase in size by about two 
lattice constants within $1.5\,t_c$ after the reactions begin, and then stay almost 
constant. 
In the meantime, the total population number of species $A$ decreases exponentially 
as displayed in Fig.~\ref{fig3}, which indicates that the number of predator $A$ 
clusters decreases quite fast. 
Fig.~\ref{fig4}(c) does not show prominent changes for the values of $\xi_{AB}(t)$ 
as the reaction time $t$ increases, demonstrating that species $A$ and $B$ maintain 
a roughly constant distance throughout the simulation.
In contrast, Fig.~\ref{fig4}(d) depicts a significant temporal evolution of 
$\xi_{BB}(t)$: the values of $\xi_{BB}$ are initially close to those of $\xi_{AA}$, 
because of the coevolution of both predator species $A$ and $B$; after several decay 
times $t_c$, however, there are few predator $A$ particles left in the system. 
The four curves for $\xi_{BB}$ would asymptotically converge after species $A$ has 
gone fully extinct.  

To summarize this section, the two indirectly competing predator species cannot 
coexist in the lattice three-species model with fixed reaction rates. 
The characteristic time for the weaker predator species to go extinct diverges as 
its reaction rates approach those of the stronger species. 
We do not observe large fluctuations of the correlation lengths during the system's 
time evolution, indicating that spatial structures remain quite stable throughout 
the Monte Carlo simulation.

\section{Introducing character displacement}
\label{subsec3}

\subsection{Model description}
\label{subsec3.1}

The Lotka--Volterra model simply treats the individuals in each population as 
particles endowed with uniform birth, death, and predation rates.  
This does not reflect a natural environment where organisms from the same species 
may still vary in predation efficiency and death or reproduction rates because of 
their size, strength, age, affliction with disease, etc. 
In order to describe individually varying efficacies, we introduce a new character 
$\eta\in[0,1]$, which plays the role of an effective trait that encapsulates the 
effects of phenotypic changes and behavior on the predation / evasion capabilities,
assigned to each individual particle \cite{uu2013}. 
When a predator $A_i$ (or $B_j$) and a prey $C_k$ occupy neighboring lattice sites,
we set the probability $(\eta_{Ai} + \eta_{Ck})/2$ [or $(\eta_{Bj} + \eta_{Ck})/2$]
for $C_k$ to be replaced by an offspring predator $A_z$ (or $B_z$). 
The indices $i$, $j$, $k$, and $z$ here indicate specific particles from the 
predator populations $A$ or $B$, the prey population $C$, and the newly created 
predator offspring in either the $A$ or $B$ population, respectively.
In order to confine all reaction probabilities in the range $[0,1]$, the efficiency 
$\eta_{Az}$ (or $\eta_{Bz}$) of this new particle is generated from a truncated 
Gaussian distribution that is centered at its parent particle efficiency $\eta_{Ai}$
(or $\eta_{Bj}$) and restricted to the interval $[0,1]$, with a certain prescribed 
distribution width (standard deviation) $\omega_{\eta A}$ (or $\omega_{\eta B}$). 
When a parent prey individual $C_i$ gives birth to a new offspring particle $C_z$, 
the efficiency $\eta_{Cz}$ is generated through a similar scheme with a given width 
$\omega_{\eta C}$. 
Thus any offspring's efficiency entails inheriting its parent's efficacy but with 
some random mutational adaptation or differentiation. 
The distribution width $\omega$ models the potential range of the evolutionary trait
change: for larger $\omega$, an offspring's efficiency is more likely to differ from 
its parent particle. 
Note that the width parameters $\omega$ here are unique for particles from the same 
species, but may certainly vary between different species. 
In previous work, we studied a two-species system (one predator and one prey) with 
such demographic variability \cite{uu2013, uu022013}. 
In that case, the system arrived at a final steady state with stable stationary
positive species abundances.
On a much faster time scale than the species density relaxation, their respective 
efficiency $\eta$ distributions optimized in this evolutionary dynamics, namely: 
the predators' efficacies rather quickly settled at a distribution centered at values 
near~$1$, while the prey efficiencies tended to small values close to $0$. 
This represents a coevolution process wherein the predator population on average 
gains skill in predation, while simultaneously the prey become more efficient in
evasion so as to avoid being killed. 

\subsection{Quasi-species mean-field equations and numerical solution}
\label{subsec3.2}

We aim to construct a mean-field description in terms of quasi-subspecies that are
characterized by their predation efficacies $\eta$.
To this end, we discretize the continuous interval of possible efficiencies
$0\leq \eta\leq 1$ into $N$ bins, with the bin midpoint values $\eta_i = (i+1/2)/N$,
$i=0, \ldots, N-1$. 
We then consider a predator (or prey) particle with an efficacy value in the range
$\eta_i-1/2\leq \eta \leq \eta_i+1/2$ to belong to the predator (or prey) subspecies
$i$. 
The probability that an individual of species $A$ with predation efficiency $\eta_1$ 
produces offspring with efficiency $\eta_2$ is assigned by means of a reproduction 
probability function $f(\eta_1, \eta_2)$. 
In the binned version, we may use the discretized form $f_{ij}=f(\eta_i, \eta_j)$.
Similarly, we have a reproduction probability function $g_{ij}$ for predator species
$B$ and $h_{ij}$ for the prey $C$.
Finally, we assign the arithmetic mean $\lambda_{ik}=(\eta_i+\eta_k)/2$ to set the 
effective predation interaction rate of predator $i$ with prey $k$ 
\cite{uu2013, uu022013}. 

These prescriptions allow us to construct the following coupled mean-field rate
equations for the temporal evolution of the subspecies populations:
\begin{equation}
\begin{split}
\label{lvreac32}
	\frac{\partial a_i(t)}{\partial t} 
		&= -\mu a_i(t) + \sum_{jk}\lambda_{kj}f_{ki}a_k(t)c_j(t)\,, \\
	\frac{\partial b_i(t)}{\partial t} 
		&= -\mu b_i(t) + \sum_{jk}\lambda_{kj}g_{ki}b_k(t)c_j(t)\,, \\
	\frac{\partial c_i(t)}{\partial t} &= \sigma\sum_kh_{ki}c_k(t)
		\bigg(1-\frac{\sum_z[a_z(t)+b_z(t)+c_z(t)]}{K}\bigg)\\
		&\quad -\sum_j\lambda_{ji}a_j(t)c_i(t) - \sum_j\lambda_{ji}b_j(t)c_i(t)\,.
\end{split}
\end{equation}
Steady-state solutions are determined by setting the time derivatives to zero, 
$\partial a_i(t)/\partial t = \partial b_i(t)/\partial t =\partial c_i(t)/\partial t
 = 0$.
Therefore, the steady-state particle counts can always be found by numerically 
solving the coupled implicit equations
\begin{equation}
\begin{split}
\label{lvreac322}
	 \mu a_i &= \sum_{jk}\lambda_{kj}f_{ki}a_kc_j\,, \\
	 \mu b_i &= \sum_{jk}\lambda_{kj}g_{ki}b_kc_j\,, \\
	 \sigma\sum_kh_{ki}c_k(t)\bigg(1-\frac{\sum_z[a_z(t)+b_z(t)+c_z(t)]}{K}\bigg) 
             &= \sum_j\lambda_{ji}a_jc_i+ \sum_j\lambda_{ji}b_jc_i\,.
\end{split}
\end{equation}
In the special case of a uniform inheritance distribution for all three species, 
$f_{ij}=g_{ij}=h_{ij}=1/N$, the above equations can be rewritten as
\begin{equation}
\begin{split}
\label{lvreac323}
	 \mu (a_i+b_i) &= \frac{1}{N}\sum_{jk}\lambda_{kj}(a_k+b_k)c_j\,, \\
	 \frac{1}{N}\sigma\sum_kc_k\bigg(1-\frac{\sum_z[a_z(t)+b_z(t)+c_z(t)]}{K}\bigg) 
                   &= \sum_j\lambda_{ji}(a_j+b_j)c_i\,,
\end{split}
\end{equation}
whose non-zero solutions are 
\begin{equation}
\label{lvreac324}
\begin{split}
&(i)\quad a_i=0, \quad \frac{b_i}{\sum_jb_j}=\frac{1}{N}\,, \quad
\frac{c_i}{\sum_jc_j}=\frac{2}{N\ln3}\frac{1}{1+2\eta_i}\,;  \\
&(ii)\quad b_i=0, \quad \frac{a_i}{\sum_ja_j}=\frac{1}{N}\,, \quad
\frac{c_i}{\sum_jc_j}=\frac{2}{N\ln3}\frac{1}{1+2\eta_i}\,;  \\
&(iii)\quad \frac{a_i+b_i}{\sum_j(a_j+b_j)}=\frac{1}{N}\,, \quad
\frac{c_i}{\sum_jc_j}=\frac{2}{N\ln3}\frac{1}{1+2\eta_i}\,. 
\end{split}
\end{equation}
We could not obtain the full time-dependent solutions to the mean-field equations
in closed form. 
We therefore employed an explicit fourth-order Runge--Kutta scheme to numerically 
solve eqs.~\eqref{lvreac32}, using a time step of $\Delta t=0.1$, the initial 
condition $a_i(t=0)=b_i(t=0)=c_i(t=0)=1/(3N)$ for $i=1, ..., N$, a number of 
subspecies $N=100$, and the carrying capacity $K=1$. 
An example for the resulting time evolution of the predator $B$ density is shown 
in Fig.~\ref{fig5}(b); its caption provides the remaining parameter values.

\subsection{Lattice simulation}
\label{subsec3.3}

We now proceed to Monte Carlo simulations for this system on a two-dimensional 
square lattice, and first study the case where trait evolution is solely 
introduced to the predation efficiencies $\eta$. 
In these simulation, the values of $\mu$ and $\sigma$ are held fixed, as is the 
nonzero distribution width $\omega$, so that an offspring's efficiency usually 
differs from its parent particle. 
In accord with the numerical solutions for the mean-field equations 
\eqref{lvreac32}, we find that the three-species system (predators $A$ and $B$, 
prey $C$) is generically unstable and will evolve into a final two-species steady 
state, where one of the predator species goes extinct, depending only on the value
of $\omega$ (given that $\mu$ and $\sigma$ are fixed). 

At the beginning of the simulation runs, the initial population densities, which 
are the particle numbers of each species divided by the lattice site number, are 
assigned the same value $0.3$ for all the three species. 
The particles are randomly distributed on the lattice sites. 
We have checked that the initial conditions do not influence the final state by 
varying the initial population densities and efficiencies. 
We fix the predator death rate to $\mu=0.125$ for both species $A$ and $B$, and set 
the prey reproduction rate as $\sigma=1.0$. 
The predation efficacies for all particles are initialized at $\eta=0.5$. 
We have varied the values of the distribution width $\omega$ and observed the final 
(quasi-)steady states. 
For the purpose of simplification, we fix $\omega_{\eta A}=\omega_{\sigma C}=0.1$, 
and compare the final states when various values of $\omega_{\eta B}$ are assigned. 

\begin{figure*}[t]
\centering
\includegraphics[width=0.43\columnwidth]{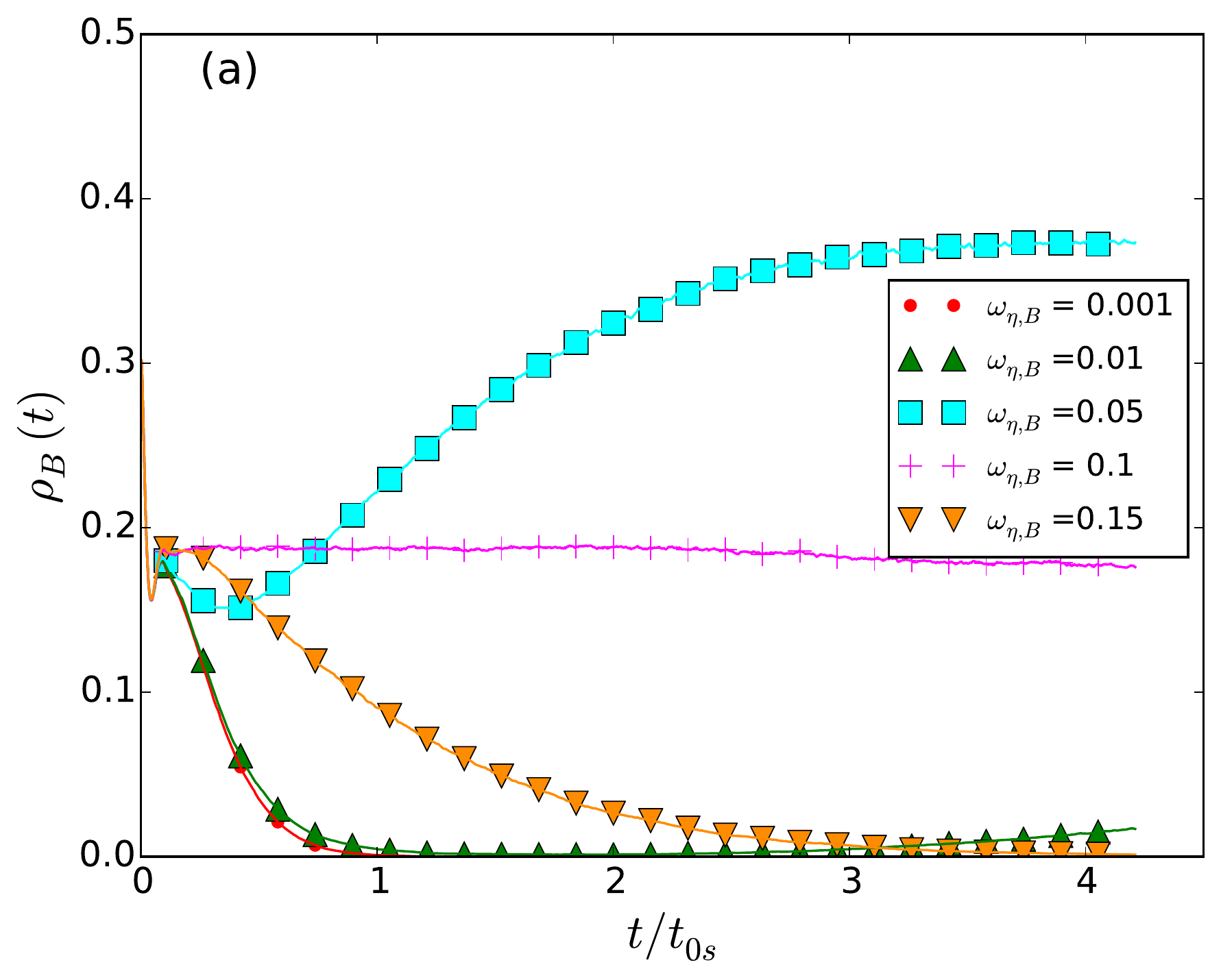} 
\includegraphics{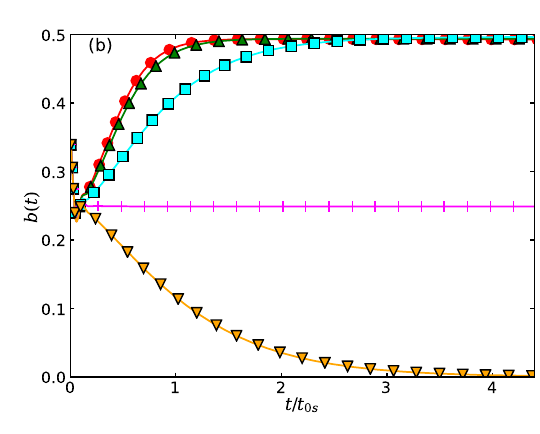}
\caption{(a) Stochastic lattice simulation of the two-predator-one-prey model with 
	`Darwinian' evolution introduced only for the predation efficiency $\eta$: 
	predator population density $\rho_B(t)$ for various values of the predation 
	efficiency distribution width $\omega_{\eta B} = 0.001$ (red dots), $0.01$ 
	(green triangles up), $0.05$ (blue squares), $0.1$ (pink crosses), and $0.15$ 
	(orange triangles down), with all other reaction rates held fixed at 
	$\mu=0.125$, $\sigma=1.0$, and $\omega_{\eta A} = \omega_{\eta C} = 0.1$; Monte
	Carlo time $t$ rescaled with the relaxation time $t_{0s}=1900$ MCS of the curve 
	for $\omega_{\eta B}=0.05$. 
	(b) Numerical solution of the mean-field eqs.~\eqref{lvreac32} with 
	$b(t)=\frac{1}{N}\sum_ib_i(t)$  denoting the average subspecies density. 
	The parameters are set at the same values as for the lattice simulations; time 
	$t$ is again normalized with the relaxation time $t_0=204.32$ of the curve for 
	$\omega_{\eta B}=0.05$ curve to allow direct comparison with the simulation 
	data. Note that the limited carrying capacity in both lattice simulations and
	the mean-field model introduces strong damping which suppresses the 
	characteristic LV oscillations.}
\label{fig5}
\end{figure*}

Fig.~\ref{fig5}(a) shows the population density $\rho_B(t)$ of predator species $B$ 
with the listed values for $\omega_{\eta B}$.
Each curve depicts a single simulation run. 
When $\omega_{\eta B} > 0.1$, the $\rho_B(t)$ quickly tends to zero; following the
extinction of the $B$ species, the system reduces to a stable $A$-$C$ two-species 
predator-prey ecology.
When $\omega_{\eta B} = 0.1$, there is no difference between species $A$ and $B$, 
so both populations survive with identical final population density; for 
$\omega_{\eta B}=0.01, 0.05$, predator species $A$ finally dies out and the system 
is reduced to a $B$-$C$ two-species system; we remark that the curve for 
$\omega_{\eta B}=0.01$ (green triangles up) decreases first and then increases again
at very late time points which is only partially shown in the graph. 
For $\omega_{\eta B}=0.001$ and even smaller, $\rho_B(t)$ goes to zero quickly, 
ultimately leaving an $A$-$C$ two-species system. 
We tried another $100$ independent runs and obtained the same results: 
for $\omega_{\eta B} \neq \omega_{\eta A}$, one of the predator species will vanish
and the remaining one coexists with the prey $C$. 
When $\omega_{\eta B}$ is smaller than $\omega_{\eta A}$ but not too close to zero, 
predator species $B$ prevails, while $A$ goes extinct. 
For $\omega_{\eta B} = 0$, there is of course no evolution for these predators at 
all, thus species $A$ will eventually outlast $B$. 
Thus there exists a critical value $\omega_{Bc}$ for the predation efficacy 
distribution width $\omega_{\eta B}$, at which the probability of either predator
species $A$ or $B$ to win the `survival game' is $50\%$. 
When $\omega_{Bc} < \omega_{\eta B} < \omega_{\eta A}$, $B$ has an advantage over 
$A$, i.e., the survival probability of $B$ is larger than $50\%$; conversely, for 
$\omega_{Bc} > \omega_{\eta B}$, species $A$ outcompetes $B$.
This means that the evolutionary `speed' is important in a finite system, and is 
determined by the species plasticity $\omega$.

Fig.~\ref{fig5}(b) shows the numerical solution of the associated mean-field 
model defined by eqs.~\eqref{lvreac32}. 
In contrast to the lattice simulations, small $\omega_{\eta B}$ do not yield 
extinction of species $B$; this supports the notion that the reentrant phase 
transition from $B$ to $A$ survival at very small values of $\omega_{\eta B}$ is 
probably a finite-size effect, as discussed below. 
Because of the non-zero carrying capacity, oscillations of population densities are 
largely suppressed in both Monte Carlo simulations and the mean-field model.
Spatio-temporal correlations in the stochastic lattice system rescale the reaction 
rates, and induce a slight difference between the steady-state population densities
in Figs.~\ref{fig5}(a) and (b) even though the microscopic rate parameters are set 
to identical values. 
For example, for $\omega_{\eta B}=0.1$, the quasi-stationary population density of 
predator species $B$ is $\approx 0.19$ (pink plus symbols) in the lattice model, 
but reaches $0.25$ in the numerical solution of the mean-field rate equations. 
Time $t$ is measured in units of Monte Carlo Steps (MCS) in the simulation; there 
is no method to directly convert this (discrete) Monte Carlo time to the continuous
time in the mean-field model. 
For the purpose of comparing the decay of population densities, we therefore 
normalize time $t$ by the associated relaxation times $t_{0s}=1900$ MCS in the 
simulations and $t_0=204.32$ in the numerical mean-field solution; both are 
calculated by performing a Fourier transform of the time-dependent prey 
densities $\rho_B(t)$ and $b(t)$ for $\omega_{\eta B}=0.05$ (blue squares).  
 
\begin{figure*}[t]
\centering
\includegraphics[width=0.7\columnwidth]{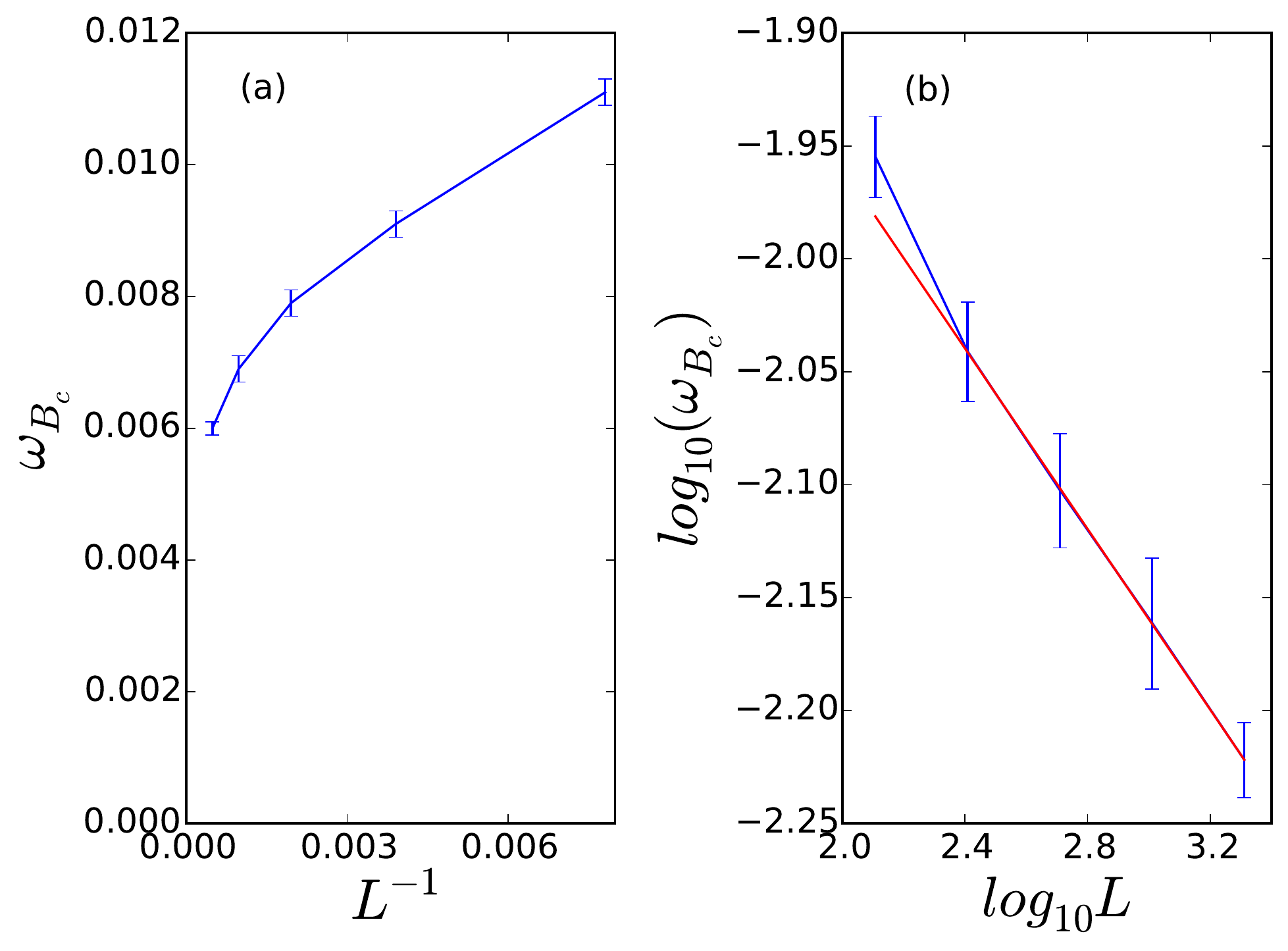} 
\caption{(a) Stochastic lattice simulation of the two-predator-one-prey model with 
	`Darwinian' evolution only introduced to the predation efficiencies $\eta$: 
	critical distribution width $\omega_{Bc}$ as a function of the inverse linear
	system size $1/L$, with predator death rate $\mu=0.125$, prey reproduction 
	rate $\sigma=1.0$, and $\omega_{\eta A} = 0.1$. 
	The data are obtained for linear system sizes 
	$L\in [128, 256, 512, 1024, 2048]$.
	(b) Double-logarithmic plot of the critical width $\omega_{Bc}$ as a function 
	of system size $L$; the red straight line represents a simple linear regression
	of the four points with $L\in[256, 512, 1024, 2048]$, with slope $-0.2$. 
	The point with $L=128$ presumably deviates from this straight line due to  
	additional strong finite-size effects.}
\label{fig6}
\end{figure*}

Our method to estimate $\omega_{Bc}$ was to scan the value space of 
$\omega_{\eta B}\in[0, 1]$, and perform $1000$ independent simulation runs for each 
value until we found the location in this interval where the survival 
probability for either $A$ or $B$ predator species was $50\%$. 
With the simulations on a $512 \times 512$ system and all the parameters set as 
mentioned above, $\omega_{Bc}$ was measured to be close to $0.008$. 
We repeated these measurements for various linear system sizes $L$ in the range 
$[128, 2048]$. 
Fig.~\ref{fig6}(a) shows $\omega_{Bc}$ as a function of $1/L$, indicating that 
$\omega_{Bc}$ decreases with a divergent rate as the system is enlarged. 
Because of limited computational resources, we were unable to extend these results 
to even larger systems. 
According to the double-logarithmic analysis shown in Fig.~\ref{fig6}(b), we 
presume that $\omega_{Bc}$ would fit a power law $\omega_{Bc}\sim L^{-\theta}$ with
exponent $\theta=0.2$. 
This analysis suggests that $\omega_{Bc}=0$ in an infinitely large system, and that
the reentrant transition from $B$ survival to $A$ survival in the range 
$\omega_{\eta B}\in [0, \omega_{\eta A}]$ is likely a finite-size effect. 
We furthermore conclude that in the three-species system (two predators and a single
prey) the predator species with a smaller value of the efficiency distribution width 
$\omega$ always outlives the other one. 
A smaller $\omega$ means that the offspring's efficiency is more centralized at its 
parent's efficacy; mutations and adaptations have smaller effects. 
Evolution may thus optimize the overall population efficiency to higher values and 
render this predator species stronger than the other one with larger $\omega$, which
is subject to more, probably deleterious, mutations. 
These results were all obtained from the measurements with $\omega_{\eta A}=0.1$.
However, other values of $\omega_{\eta A}$ including $0.2$, $0.3$, and $0.4$ were 
tested as well, and similar results observed.

Our numerical observation that two predator species cannot coexist contradicts 
observations in real ecological systems. 
This raises the challenge to explain multi-predator-species coexistence. 
Notice that `Darwinian' evolution was only applied to the predation efficiency in 
our model. 
However, natural selection could also cause lower predator death rates and increased 
prey reproduction rates so that their survival chances would be enhanced in the 
natural selection competition. 
One ecological example are island lizards that benefit from decreased body size 
because large individuals will attract attacks from their competitors \cite{j2002}. 
In the following, we adjust our model so that the other two reaction rates $\mu$ and 
$\sigma$ do not stay fixed anymore, but instead evolve by following the same 
mechanism as previously implemented for the predation efficacies $\eta$. 
The death rate of an offspring predator particle is hence generated from a truncated 
Gaussian distribution centered at its parent's value, with positive standard 
deviations  $\omega_{\mu A}$ and $\omega_{\mu B}$ for species $A$ and $B$, 
respectively. 
The (truncated) Gaussian distribution width for the prey reproduction rate is 
likewise set to a non-zero value $\omega_\sigma$. 

\begin{figure*}[]
\centering
\includegraphics[width=0.7\columnwidth]{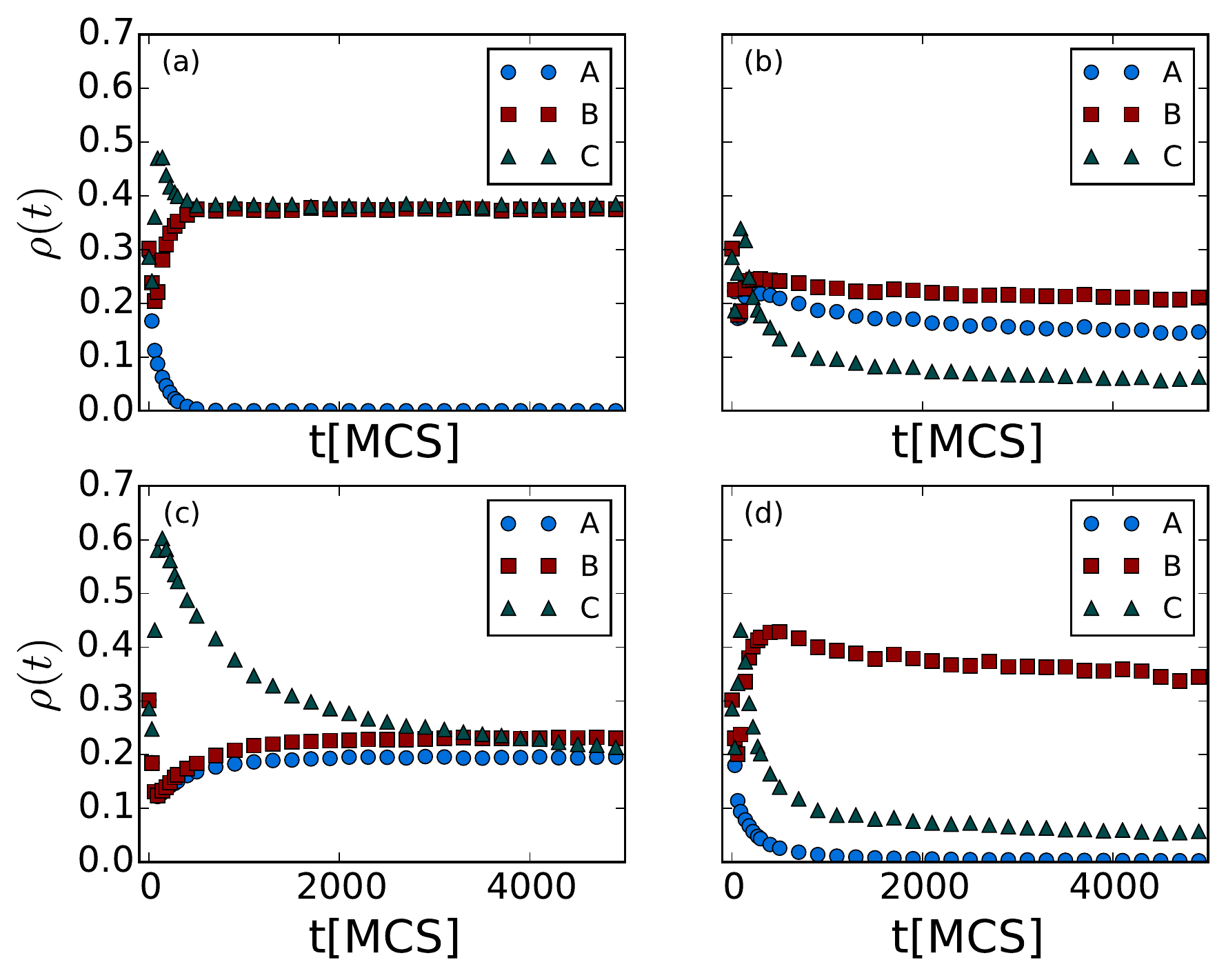} 
\caption{Population densities $\rho(t)$ from Monte Carlo simulations with 
	`Darwinian' evolution introduced to both the predation efficiencies $\eta$ and 
	predator death rates $\mu$, while the prey reproduction rate stays fixed at 
	$\sigma=1.0$. 
	The species are indicated as blue dots for $A$, red squares for $B$, and green 
	triangles for $C$. 
	The final states are in (a) $A$ extinction; (b) and (c) transient three-species
	coexistence; and (d) $B$ extinction, with
	$\omega_{\eta A}=0.11$, $\omega_{\mu A}=0.3$, $\omega_{\mu B}=0.125$ in (a), 
	$\omega_{\eta A}=0.08$, $\omega_{\mu A}=0.1$, $\omega_{\mu B}=0.09$ in (b), 
	$\omega_{\eta A}=0.08$, $\omega_{\mu A}=0.4$, $\omega_{\mu B}=0.39$ in (c), and
	$\omega_{\eta A}=0.08$, $\omega_{\mu A}=0.4$, $\omega_{\mu B}=0.09$ in (d), 
	while $\omega_{\eta B}=0.1$, $\omega_{\eta C}=0.1$, $\omega_{\sigma C}=0$ for 
	all four plots.}
\label{fig7}
\end{figure*}

In the simulations, the initial population densities for all three species are set 
at $0.3$ with the particles randomly distributed on the lattice. 
The reaction rates and efficiencies for these first-generation individuals were 
chosen as $\eta_{A0}=\eta_{B0}=\eta_{C0}=0.5$, $\mu_{A0}=\mu_{B0}=0.125$, and 
$\sigma_0=1.0$. 
With this same initial set, we ran simulations with different values of the 
Gaussian distribution widths $\omega$. 
Figure~\ref{fig7} displays the temporal evolution of the three species' population 
densities with four sets of given widths $\omega$: 
In Fig.~\ref{fig7}(a), $\omega_{\eta A}=0.11$, $\omega_{\eta B}=0.1$, 
$\omega_{\eta C}=0.1$, $\omega_{\mu A}=0.3$, $\omega_{\mu B}=0.125$, and
$\omega_{\sigma C}=0$. 
Since a smaller width $\omega$ gives advantages to the corresponding species, 
$\omega_{\eta B} < \omega_{\eta A}$ and $\omega_{\mu B} < \omega_{\mu A}$ render 
predators $B$ stronger than $A$ in general. 
As the graph shows, species $A$ dies out quickly and finally only $B$ and $C$ 
remain in the system. 
In all four cases, the prey $C$ stay active and do not become extinct.  

However, it is not common that a species is stronger than others in every aspect, 
so we next set $\omega$ so that $A$ has advantages over $B$ in predation, i.e., 
$\omega_{\eta A} < \omega_{\eta B}$, but is disadvantaged through 
broader-distributed death rates $\omega_{\mu A} > \omega_{\mu B}$. 
In Fig.~\ref{fig7}(b), $\omega_{\eta A}=0.08$, $\omega_{\eta B}=0.1$, 
$\omega_{\eta C}=0.1$, $\omega_{\mu A}=0.1$, $\omega_{\mu B}=0.09$, and
$\omega_{\sigma C}=0$; in Fig.~\ref{fig7}(c), $\omega_{\eta A}=0.08$, 
$\omega_{\eta B}=0.1$, $\omega_{\eta C}=0.1$, $\omega_{\mu A}=0.4$, 
$\omega_{\mu B}=0.39$, and $\omega_{\sigma C}=0$. 
In either case, none of the three species becomes extinct after $10\,000$ MCS, and 
three-species coexistence will persist at least for much longer time. 
Monitoring the system's activity, we see that the system remains in a dynamic state
with a large amount of reactions happening. 
When we repeat the measurements with other independent runs, similar results are 
observed, and we find the slow decay of the population densities to be rather 
insensitive to the specific values of the widths $\omega$. 
As long as we implement a smaller width $\omega$ for the $A$ predation efficiency 
than for the $B$ species, but a larger one for its death rates, or vice versa, 
three-species coexistence emerges.
Of course, when the values of the standard deviations $\omega$ differ too much 
between the two predator species, one of them may still approach extinction fast. 
One example is shown in Fig.~\ref{fig7}(d), where $\omega_{\eta A}=0.08$, 
$\omega_{\eta B}=0.1$, $\omega_{\eta C}=0.1$, $\omega_{\mu A}=0.4$, 
$\omega_{\mu B}=0.09$, and $\omega_{\sigma C}=0$; since $\omega_{\mu A}$ is about 
five times larger than $\omega_{\mu B}$ here, the predation advantage of species
$A$ cannot balance its death rate disadvantage, and consequently species $A$ is 
driven to extinction quickly. 
Yet the coexistence of all three competing species in Figs.~\ref{fig7}(b) and (c) 
does not persist forever, and at least one species will die out eventually, after an extremely long time. 
Within an intermediate time period, which still amounts to thousands of generations, 
they can be regarded as quasi-stable because the decay is very slow. 
This may support the idea that in real ecosystems perhaps no truly stable 
multiple-species coexistence exists, and instead the competing species are in fact 
under slow decay which is not noticeable within much shorter time intervals. 
In Figs.~\ref{fig7}(a) and (d), the predator $A$ population densities decay 
exponentially with relaxation times of order $100$ MCS, while the corresponding
curves in (b) and (c) approximately follow algebraic functions (power law decay).

However, we note that in the above model implementation the range of predator death 
rates $\mu$ was the entire interval $[0,1]$, which gives some individuals a very low
chance to decay. 
Hence these particles will stay in the system for a long time, which accounts for 
the long-lived transient two-predator coexistence regime. 
To verify this assumption, we set a positive lower bound on the predators' death 
rates, preventing the presence of near-immortal individuals.
We chose the value of the lower bound to be $0.001$, with the death rates $\mu$ for
either predator species generated in the predation and reproduction processes having
to exceed this value. 
Indeed, we observed no stable three-species coexistence state, i.e., one of the 
predator species was invariably driven to extinction, independent of the values of 
the widths $\omega$, provided they were not exactly the same for the two predator 
species. 
To conclude, upon introducing a lower bound for their death rates, the two predator 
species cannot coexist despite their dynamical evolutionary optimization.

\section{Effects of direct competition between both predator species}
\label{sec4}

\subsection{Inclusion of direct predator competition and mean-field analysis}
\label{subsec4.1}

We proceed to include explicit direct competition between both predator species in 
our model.
The efficiencies of predator particles are most likely to be different since they 
are randomly generated from truncated Gaussian distributions. 
When a strong $A$ individual (i.e., with a large predation efficacy $\eta$) meets a 
weaker $B$ particle on an adjacent lattice site, or vice versa, we now allow 
predation between both predators to occur. 
Direct competition is common within predator species in nature. 
For example, a strong lizard may attack and even kill a small lizard to occupy its 
habitat. 
A lion may kill a wolf, but an adult wolf might kill an infant lion.  
Even though cannibalism occurs in nature as well, we here only consider direct 
competition and predation between different predator species. 
In our model, direct competition between the predator species is implemented as 
follows: 
For a pair of predators $A_i$ and $B_j$ located on neighboring lattice sites and 
endowed with respective predation efficiencies $\eta_{Ai}$ and 
$\eta_{Bj} < \eta_{Ai}$, particle $B_j$ is replaced by a new $A$ particle $A_z$ 
with probability $\eta_{Ai} - \eta_{Bj}$; conversely, if $\eta_{Ai} < \eta_{Bj}$, 
there is a probability $\eta_{Bj} - \eta_{Ai}$ that $A_i$ is replaced by a new 
particle $B_z$.

We first write down and analyze the mean-field rate equations for the simpler case 
when the predator species compete directly without evolution, i.e., all reaction 
rates are uniform and constant. 
We assume that $A$ is the stronger predator with $\lambda_A > \lambda_B$, hence 
only the reaction $A+B\to A+A$ is allowed to take place with rate 
$\lambda_A-\lambda_B$, but not its complement, supplementing the original reaction 
scheme listed in \eqref{lvreac21}. 
The associated mean-field rate equations read
\begin{equation}
\label{lvreac42}
\begin{split}
	\frac{d a(t)}{dt} &= -\mu_A a(t) + \lambda_A a(t)c(t) 
		+ (\lambda_A-\lambda_B)a(t)b(t)\,, \\
	\frac{d b(t)}{dt} &= -\mu_B b(t) + \lambda_B b(t)c(t) 
		- (\lambda_A-\lambda_B)a(t)b(t)\,, \\
	\frac{d c(t)}{dt} &= \sigma c(t)\bigg[1-\frac{a(t)+b(t)+c(t)}{K}\bigg] 
		- \lambda_A a(t)c(t) -\lambda_B b(t)c(t)\,,
\end{split}
\end{equation}
with the non-zero stationary solutions
\begin{equation}
\label{lvreac420}
\begin{split}
	(i)\quad a=0\,,\quad 
	b=\frac{\sigma(K\lambda_B-\mu_B)}{\lambda_B(\sigma+K\lambda_B)}\,,\quad 
	c=\frac{\mu_B}{\lambda_B}\,,\\
	(ii)\quad a=\frac{\sigma(K\lambda_A-\mu_A)}{\lambda_A(\sigma+K\lambda_A)}\,,
	\quad b=0\,,\quad c=\frac{\mu_A}{\lambda_A}\,,\\
	(iii)\quad a+b+c = \frac{\mu_A-\mu_B}{\lambda_A-\lambda_B}\,,\quad \text{when}
	\quad a(0)+b(0)+c(0) = \frac{\mu_A-\mu_B}{\lambda_A-\lambda_B}\,.
\end{split}
\end{equation}
Within this mean-field theory, three-species coexistence states exist only when the 
total initial population density is set to
$a(0)+b(0)+c(0) = \frac{\mu_A-\mu_B}{\lambda_A-\lambda_B}$. 
In our lattice simulations, however, we could not observe any three-species 
coexistence state even when we carefully tuned one reaction rate with all others 
held fixed.

\begin{figure*}[t]
\centering
\includegraphics{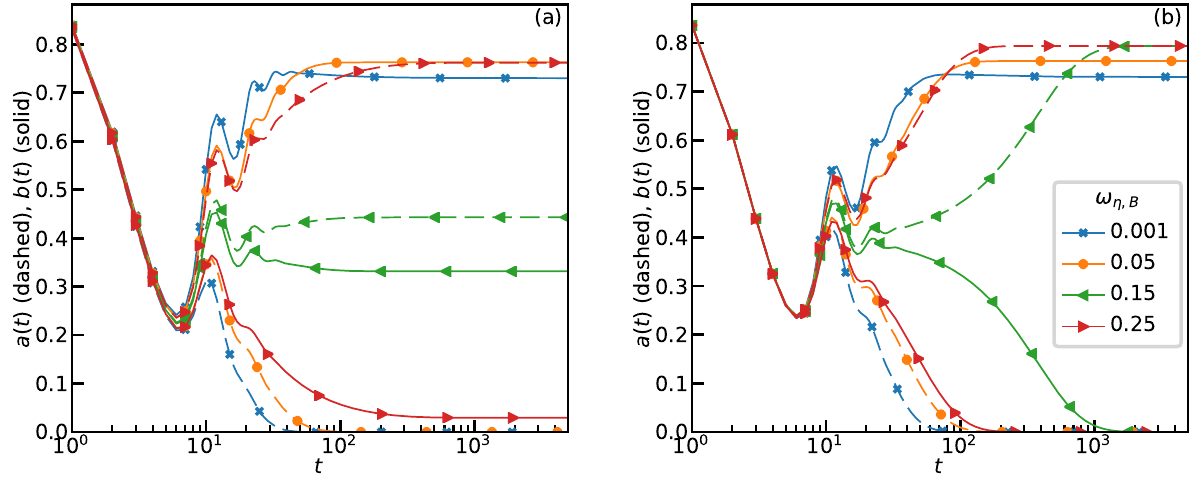} 
\caption{Numerical solutions of the mean-field equations~\eqref{eq:4.2mfequations}
	for the two-predator subspecies densities $a(t)=\frac{1}{N}\sum_ia_i(t)$ 
	(dashed) and $b(t)=\frac{1}{N}\sum_ib_i(t)$ (solid) for different distribution
	widths $\omega_{\eta,B}$ and the parameters $\omega_{\eta,A}=0.14$, 
	$\omega_{\eta,C}=\infty$, $\sigma=1$, and $\mu=0.5$. 
	(a) Population densities in the presence of direct predator-predator 
	competition; and (b) in the absence of this competition. 
	Note that three-species coexistence is only possible when direct 
	predator-predator competition is explicitly implemented.}
\label{fig9}
\end{figure*}

Next we reinstate `Darwinian' evolution for this extended model with direct 
competition between the predator species. 
We utilize the function $\hat\lambda_{ij}=|\eta_i-\eta_j|$ to define the reaction 
rate between predators $A$ and $B$.
For the case that the predator death rate $\mu$ is fixed for both species $A$ and 
$B$, the ensuing quasi-subspecies mean-field equations are
\begin{equation}
\label{eq:4.2mfequations}
\begin{split}
\frac{\partial a_i(t)}{\partial t}&=-\mu a_i(t)
	+\sum_{jk}\lambda_{kj}f_{ki}a_k(t)c_j(t)
	+\sum_{j<k}\hat\lambda_{kj}f_{ki}a_k(t)b_j(t)\\
	&\quad -\sum_{j>i}\hat\lambda_{ij}a_i(t)b_j(t)\,,\\
	\frac{\partial b_i(t)}{\partial t}&=-\mu b_i(t)
	+\sum_{jk}\lambda_{kj}g_{ki}b_k(t)c_j(t)+
	\sum_{j<k}\hat\lambda_{kj}g_{ki}b_k(t)a_j(t)\\
	&\quad -\sum_{j>i}\hat\lambda_{ji}b_i(t)a_j(t)\,\\
	\frac{\partial c_i(t)}{\partial t}&=\sigma\sum_jh_{ji}c_j(t)
	\bigg(1-\frac{\sum_z[a_z(t)+b_z(t)+c_z(t)]}{K}\bigg)\\
	&\quad -\sum_j\lambda_{ji}[a_j(t)+b_j(t)]c_i(t)\,.
\end{split}
\end{equation}
Since a closed set of solutions for eqs.~(\ref{eq:4.2mfequations}) is very 
difficult to obtain, we resort to numerical integration. 
As before, we rely on an explicit fourth-order Runge--Kutta scheme with time step
$\Delta t=0.1$, initial conditions $a_i(t=0)=b_i(t=0)=c_i(t=0)=1/N$, number of 
subspecies $N=100$, and carrying capacity $K=3$. 
Four examples for such numerical solutions of the quasi-subspecies mean-field
equations are shown in Fig.~\ref{fig9}, and will be discussed in the following 
subsection. 

\subsection{The quasi-stable three-species coexistence region}
\label{subsec4.3}

\begin{figure*}[t]
\centering
\includegraphics[width=1.0\columnwidth]{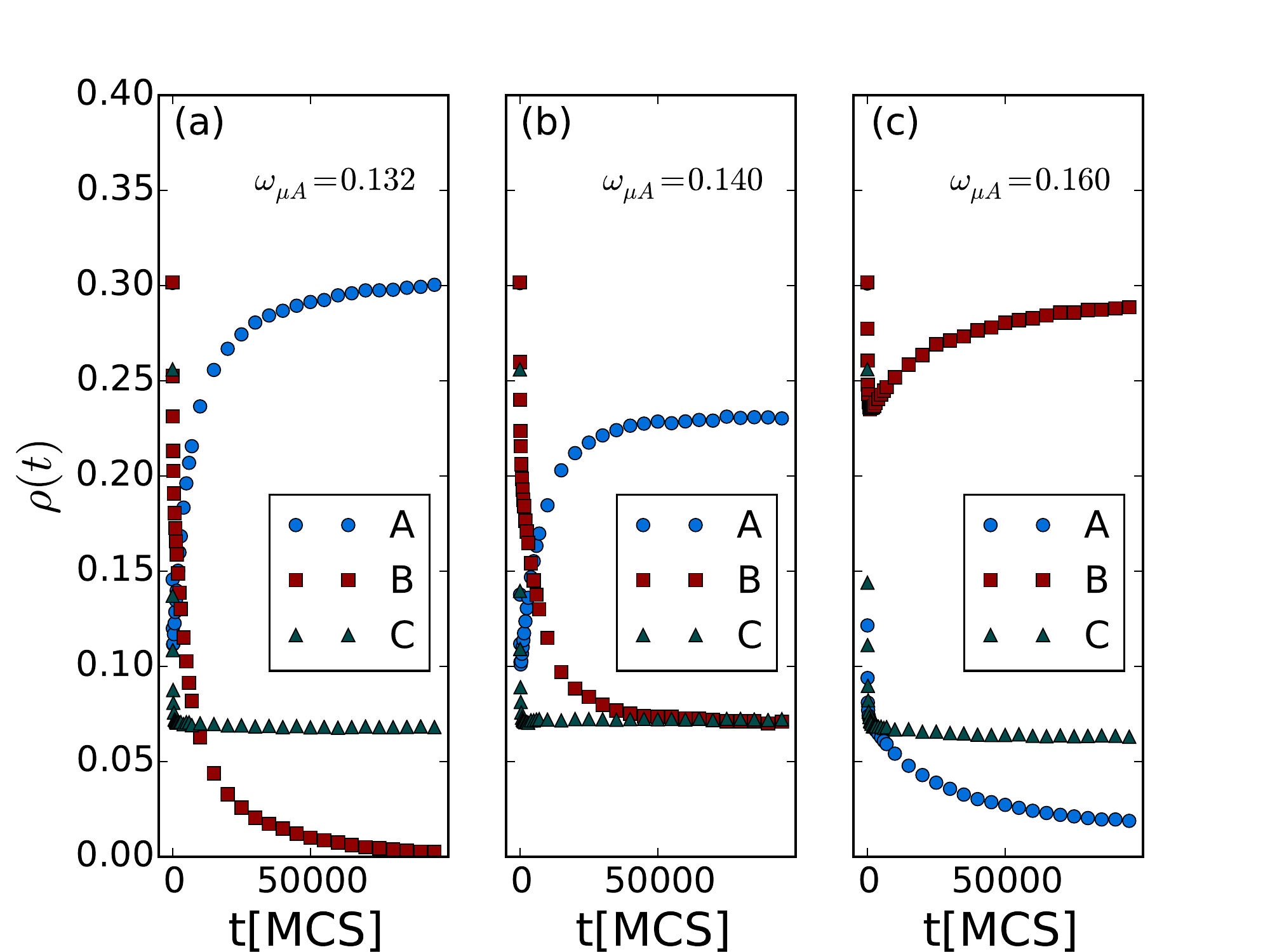} 
\caption{Data obtained from Monte Carlo simulations where both direct competition 
    between both predator species as well as evolutionary dynamics are introduced:
	Temporal population density record with $\omega_{\eta A}=0.1$, 
	$\omega_{\eta B}=0.15$, $\omega_{\mu B}=0.1$ and 
	$\omega_{\mu A}=0.132$, $0.140$, $0.160$ (from left to right) with species $A$
	indicated with blue dots, $B$ red-dashed, and $C$ with green triangles.}
\label{fig8}
\end{figure*}

For the three-species system with two predators $A$, $B$ and prey $C$, we now 
introduce `Darwinian' evolution to both the predator death rates $\mu$ and the 
predation efficiencies $\eta$. 
In addition, we implement direct competition between the predators $A$ and $B$. 
We set the lower bound of the death rates $\mu$ to $0.001$ for both predator 
species. 
The simulations are performed on a $512\times512$ square lattice with periodic 
boundary conditions. 
Initially, individuals from all three species are randomly distributed in the 
system with equal densities $0.3$. 
Their initial efficiencies are chosen as $\eta_A = 0.5 = \eta_B$ and $\eta_C = 0$.
Since there is no evolution of the prey efficiency, $\eta_C$ will stay zero 
throughout the simulation. 
The distribution widths for the predation efficiencies are fixed to 
$\omega_{\eta A}=0.1$ and $\omega_{\eta B}=0.15$, giving species $A$ an advantage
over $B$ in the non-linear predation process. 
We select the width of the death rate distribution of species $B$ as
$\omega_{\mu B}=0.1$. 
If $\omega_{\mu A}$ is also chosen to be $0.1$, the $B$ population density would 
decay exponentially. 
$\omega_{\mu A} > \omega_{\mu B}=0.1$ is required to balance species $A$'s 
predation adaptation advantage so that stable coexistence is possible. 
Figure~\ref{fig8} shows the population densities resulting from our 
individual-based Monte Carlo simulations as a function of time, for different
values $\omega_{\mu A} = 0.132$, $0.140$, and $0.160$. 
These graphs indicate the existence of phase transitions from species $B$ 
extinction in Fig.~\ref{fig8}(a) to predator $A$-$B$ coexistence in 
Fig.~\ref{fig8}(b), and finally to $A$ extinction in Fig.~\ref{fig8}(c)). 
In Fig.~\ref{fig8}(a), species $A$ is on average more efficient than $B$ in 
predation, but has higher death rates. 
Predator species $B$ is in general the weaker one, and hence goes extinct after 
about $100\,000$ MCS. 
Figure~\ref{fig8}(b) shows a (quasi-)stable coexistence state with neither 
predator species dying out within our simulation time. 
In Fig.~\ref{fig8}(c), $\omega_{\mu A}$ is set so high that $A$ particles die much
faster than $B$ individuals, so that finally species $A$ would vanish entirely.

Figure~\ref{fig9}(a) displays the time evolution for the solutions of the 
corresponding quasi-subspecies mean-field model (\ref{eq:4.2mfequations}) for four 
different values of the species $B$ efficiency width $\omega_{\eta,B}$. 
In particular, it shows that there is a region of coexistence in which both 
predator species reach a finite steady-state density, supporting the Monte Carlo
results from the stochastic lattice model. 
In contrast, numerical solutions of eqs.~(\ref{eq:4.2mfequations}) with
$\hat{\lambda}_{ij}=0$, equivalent to eqs.~(\ref{lvreac32}), exhibit no 
three-species coexistence region; see Fig.~\ref{fig9}(b). 

\begin{figure*}[t]
\centering
\includegraphics[width=0.9\columnwidth]{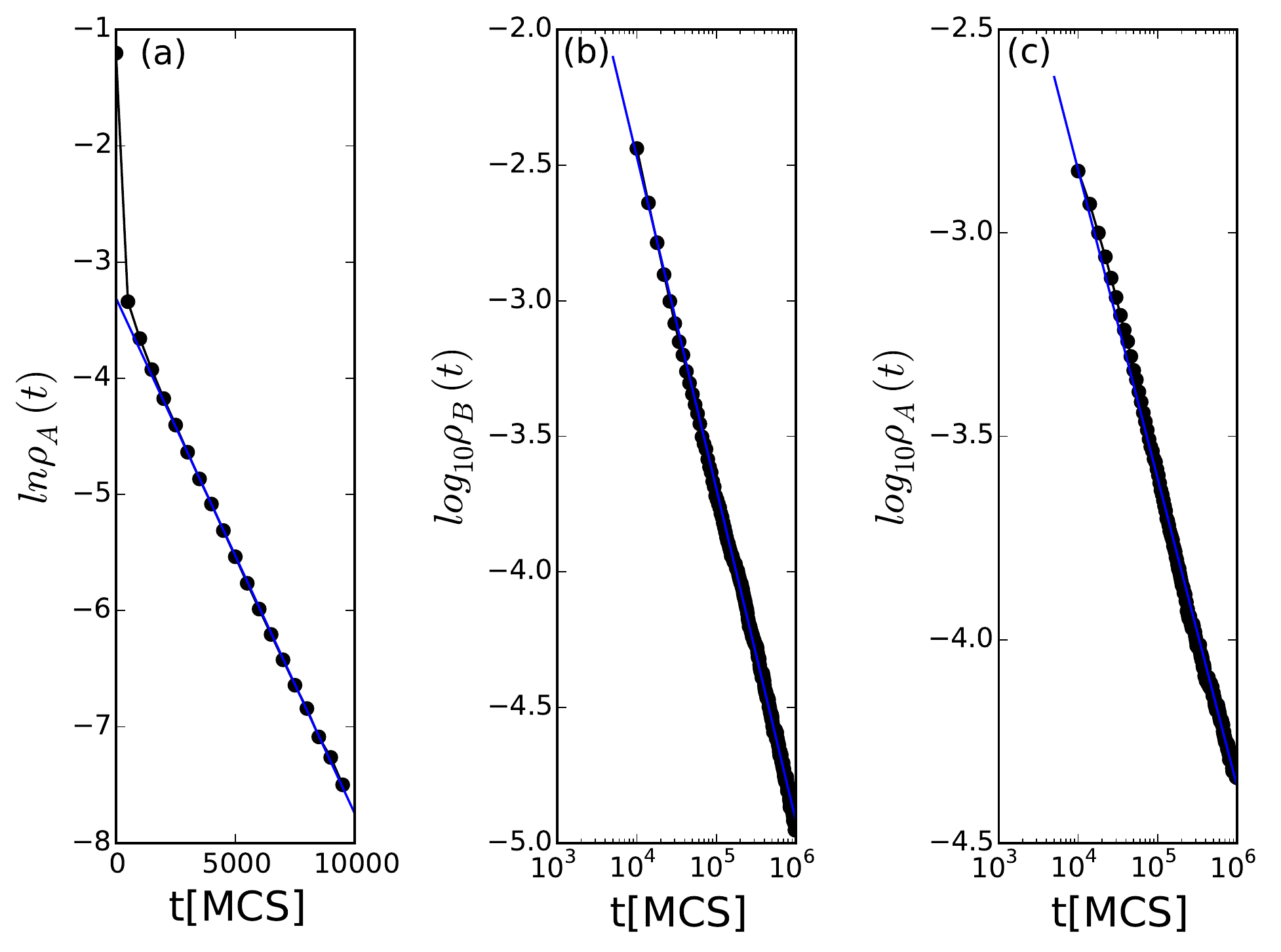} 
\caption{Monte Carlo simulations with direct predator competition:
    (a) Exponential decay of the predator population density $\rho_A(t)$ with 
	$\omega_{\mu A} = 0.2$, $\omega_{\eta A}=0.1$, $\omega_{\eta B}=0.15$, and 
	$\omega_{\mu B}=0.1$; the blue straight line is obtained from a linear
	regression of the data points for $x\geq2000$, with slope $-0.00044$.
	(b) Algebraic power-law decay of the predator $B$ species density with 
	$\omega_{\mu A} = 0.136$ and the other parameters set as in (a).
	(c) Power-law decay of $\rho_A(t)$ for $\omega_{\mu A} = 0.159$. 
	The black dots are measured population densities from the simulations, while 
	the blue straight lines indicate simple linear regressions of the simulation 
	data.}
\label{fig10}
\end{figure*}

At an active-to-absorbing phase transition threshold, one should anticipate the 
standard critical dynamics phenomenology for a continuous phase transition: 
exponential relaxation with time becomes replaced by much slower algebraic decay of
the population density \cite{noneq1, uct2014}. 
We determine the three-species coexistence range for our otherwise fixed parameter 
set to be in the range $\omega_{\mu A} \in [0.136, 0.159]$. 
Figure~\ref{fig10}(a) shows an exponential decay of the predator population $A$ 
density with $\omega_{\mu A} = 0.2$, deep in the absorbing extinction phase. 
The system would attain $B$-$C$ two-species coexistence within of the order $10^4$
MCS. 
We also ran the Monte Carlo simulation with $\omega_{\mu A} = 0.1$, also inside an 
absorbing region, but now with species $B$ going extinct, and observed exponential 
decay of $\rho_B(t)$.  
By changing the value of $\omega_{\mu A}$ to $0.136$ as plotted in 
Fig.~\ref{fig10}(b), $\rho_B(t)\sim t^{-\alpha_B}$ fits a power law decay with 
critical exponent $\alpha_B=1.22$. 
Since it would take infinite time for $\rho_B$ to reach zero while species $A$ and
$C$ densities remain finite during the entire simulation time, the system at this
point already resides at the threshold of three-species coexistence. 
Upon increasing $\omega_{\mu A}$ further, all three species densities would reach 
their asymptotic constant steady-state values within a finite time and then remain
essentially constant (with small statistical fluctuations).  
At the other boundary of this three-species coexistence region, 
$\omega_{\mu A} = 0.159$, the decay of $\rho_A(t)$ also fits a power law as 
depicted in Fig.~\ref{fig10}(c), and $\rho_B(t)$ would asymptotically reach a 
positive value. 
However, the critical power law exponent is in this case estimated to be 
$\alpha_A=0.76$. 
We do not currently have an explanation for the distinct values observed for the 
decay exponents $\alpha_A$ and $\alpha_B$, neither of which are in fact close to 
the corresponding directed-percolation value $\alpha = 0.45$ \cite{cr1997}. 
If we increase $\omega_{\mu A}$ even more, species $A$ would die out quickly and 
the system subsequently reduce to a $B$-$C$ two-species predator-prey coexistence 
state. 
We remark that the critical slowing-down of the population density at either of
the two thresholds as well as the associated critical aging scaling may serve as 
a warning signal of species extinction 
\cite{ldkj2012, su2016}.

\begin{figure*}[t]
\centering
\includegraphics[width=0.7\columnwidth]{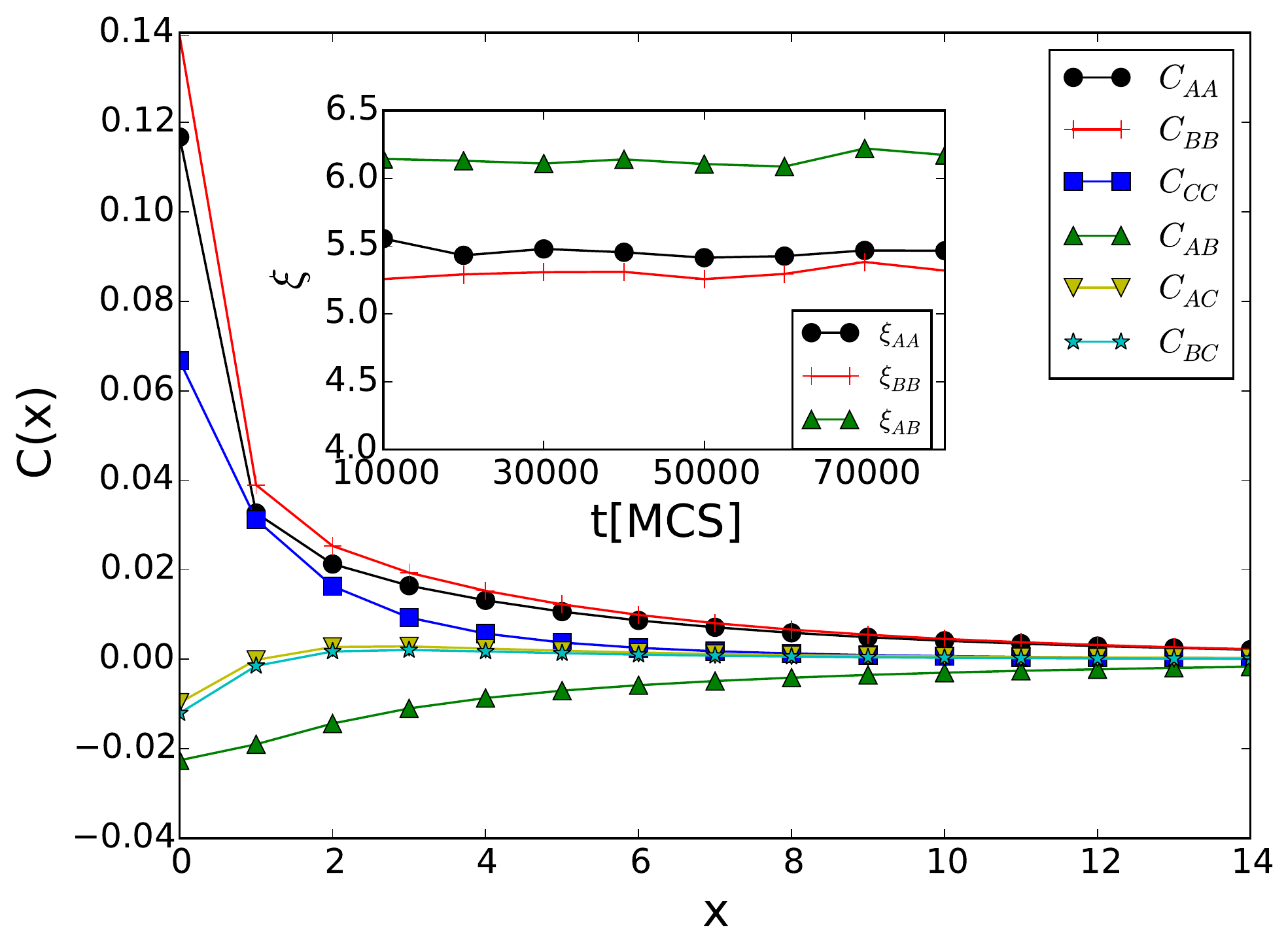} 
\caption{Monte Carlo simulations with direct predator competition.
	Main panel: Quasi-stationary correlation functions $C(x)$ after the system has
	evolved for $10\,000$ MCS with $\omega_{\mu A}=0.147$, $\omega_{\eta A}=0.1$,
	$\omega_{\eta B}=0.15$, and $\omega_{\mu B}=0.1$, when the system resides in 
	the coexistence state. 
	Inset: temporal evolution of the correlation length $\xi(t)$; all lengths are
	measured in units of the square lattice spacing.}
\label{fig11}
\end{figure*}

It is of interest to study the spatial properties of the particle distribution. 
We choose $\omega_{\mu A} = 0.147$ so that the system resides deep in the 
three-species coexistence region according to Fig.~\ref{fig10}. 
The correlation functions are measured after the system has evolved for $10\,000$
MCS as shown in the main plot of Fig.~\ref{fig11}. 
The results are similar to those in the previous sections in the sense that 
particles are positively correlated with the ones from the same species, but 
anti-correlated to individuals from other species. 
The correlation functions for both predator species are very similar: 
$C_{AA}(x)$ and $C_{BB}(x)$ overlap each other for $x\geq 5$, and $C_{AC}$ and 
$C_{BC}$ coincide for $x\geq 2$ lattice sites. 
The inset displays the measured characteristic correlation length as functions of
simulation time, each of which varies on the scale of $\sim 0.1$ during $70\,000$ 
MCS, indicating that the species clusters maintain nearly constant sizes and keep 
their respective distances almost unchanged throughout the simulations. 
The correlation lengths $\xi_{AA}$ and $\xi_{BB}$ are very close and differ only by
less than $0.2$ lattice sites. 
These data help to us to visualize the spatial distribution of the predators: 
The individuals of both $A$ and $B$ species arrange themselves in clusters with 
very similar sizes throughout the simulation, and their distances to prey clusters 
are almost the same as well. 
Hence predator species $A$ and $B$ are almost indistinguishable in their spatial 
distribution. 

\subsection{Monte Carlo simulation results in a zero-dimensional system}
\label{subsec4.4}

\begin{figure*}[t]
\centering
\includegraphics[width=1.0\columnwidth]{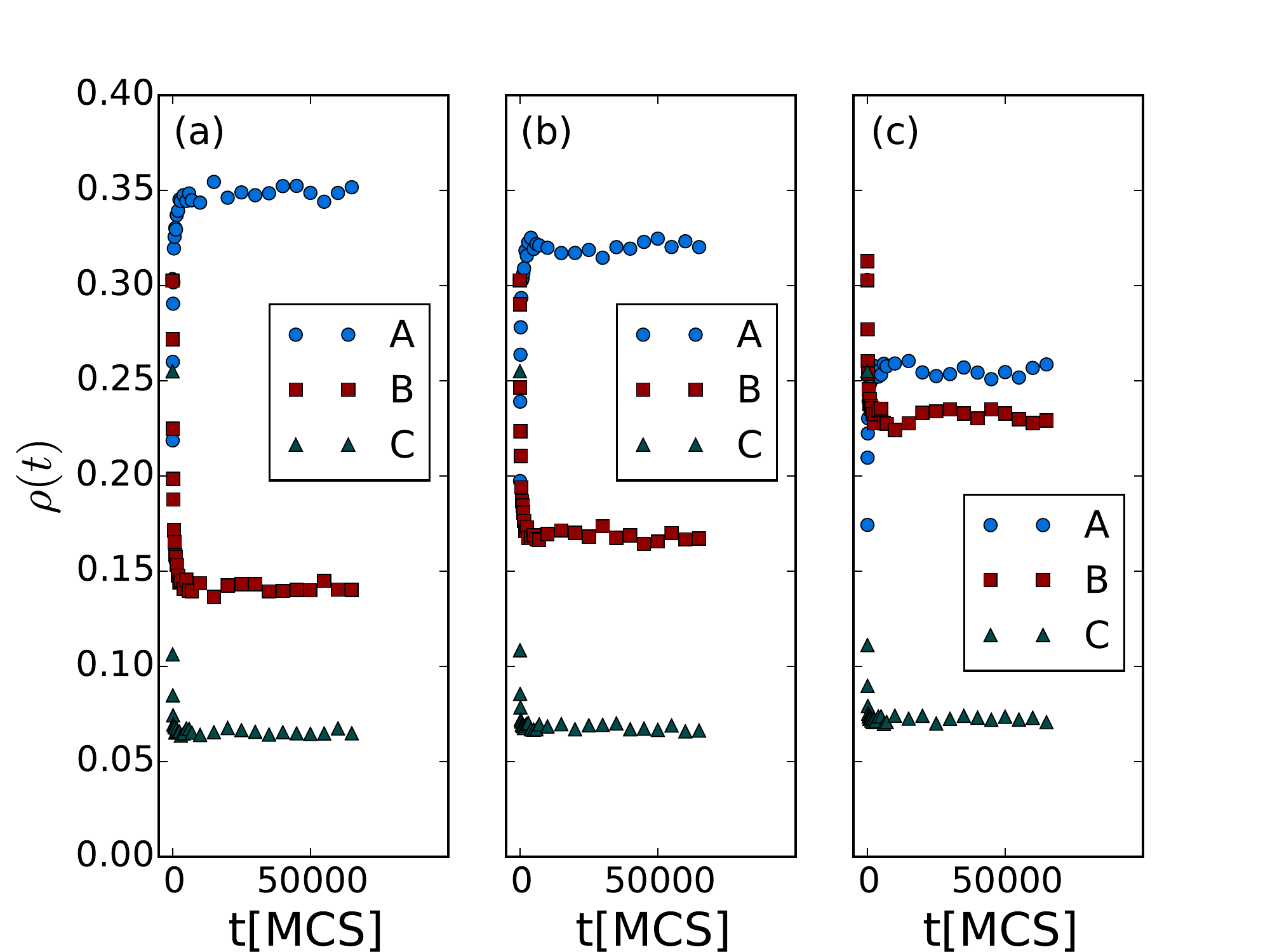} 
\caption{Data obtained from single Monte Carlo simulation runs in a 
	zero-dimensional system with direct competition and evolutionary dynamics, 
	hence only temporal but no spatial correlations:
	Time record of the population densities for all three species with 
	$\omega_{\eta A}=0.1$, $\omega_{\eta B}=0.15$, $\omega_{\mu B}=0.1$ and  
	$\omega_{\mu A}=0.132, 0.140, 0.160$ (from left to right), with species $A$ 
	indicated with blue dots, $B$ red-dashed, and $C$ with green triangles.}
\label{fig12}
\end{figure*}

The above simulations were performed on a two-dimensional system by locating the 
particles on the sites of a square lattice. 
Randomly picked particles are allowed to react (predation, reproduction) with their
nearest neighbors. 
Spatial as well as temporal correlations are thus incorporated in the reaction 
processes. 
In this subsection, we wish to compare our results with a system for which spatial
correlations are absent, yet which still displays manifest temporal correlations. 
To simulate this situation, we remove the nearest-neighbor restriction and instead 
posit all particles in a `zero-dimensional' space. 
In the resulting `urn' model, the simulation algorithm entails to randomly pick two
particles and let them react with a probability determined by their individual
character values. 
We find that if all the particles from a single species are endowed with 
homogeneous properties, i.e., the reaction rates are fixed and uniform as in 
section 2, no three-species coexistence state is ever observed. 
If evolution is added without direct competition between predator species as in 
section 3, the coexistence state does not exist neither. 
Our observation is again that coexistence occurs only when both evolution and 
direct competition are introduced.  
Qualitatively, therefore, we obtain the same scenarios as in the two-dimensional 
spatially extended system. 
The zero-dimensional system however turns out even more robust than the one on a
two-dimensional lattice, in the sense that its three-species coexistence region 
is considerably more extended in parameter space. 
Figure~\ref{fig12} displays a series of population density time evolutions from 
single zero-dimensional simulation runs with identical parameters as in 
Fig.~\ref{fig8}. 
All graphs in Fig.~\ref{fig12} reside deeply in the three-species coexistence 
region, while Fig.~\ref{fig8}(a) and (c) showed approaches to absorbing states with
one of the predator species becoming extinct.   
With $\omega_{\eta A}=0.1$, $\omega_{\eta B}=0.15$, and $\omega_{\mu B}=0.1$ fixed,
three-species coexistence states in the zero-dimensional system are found in the 
region $\omega_{\mu A} \in (0, 1)$, which is to be compared with the much 
narrower interval $(0.136, 0.159)$ in the two-dimensional system, indicating that 
spatial extent tends to destabilize these systems. 

This finding is in remarkable contrast to some already well-studied systems such as 
the three-species cyclic competition model, wherein spatial extension and disorder 
crucially help to stabilize the system \cite{uu2008, qmu2011}.
Even though we do not allow explicit nearest-neighbor `hopping' of particles in the
lattice simulation algorithm, there still emerges effective diffusion of prey 
particles followed by predators. 
Since predator individuals only have access to adjacent prey in the lattice model,
the presence of one predator species would block their neighboring predators from 
their prey. 
Imagining a cluster of predator particles surrounded by the other predator species,
they will be prevented from reaching their `food' and consequently gradually die 
out. 
However, this phenomenon cannot occur in the zero-dimensional system where no 
spatial structure exists at all, and hence blockage is absent. 
In the previous section we already observed that the cluster size of predator 
species remains almost unchanged throughout the simulation process when the total 
population size of the weaker predator species gradually decreases to zero, 
indicating that clusters vanish in a sequential way. 
We also noticed that population densities reach their quasi-stationary values much 
faster in the non-spatial model, see Fig.~\ref{fig12}, than on the two-dimensional 
lattice, Fig.~\ref{fig8}.
In the spatially extended system, particles form intra-species clusters, and 
reactions mainly occur at the boundaries between neighboring such clusters of
distinct species, thus effectively reducing the total reaction speed. 
This limiting effect is absent in the zero-dimension model where all particles have 
equal chances to meet each other.  

\subsection{Character displacements}
\label{subsec4.5}

\begin{figure*}[t]
\centering
\includegraphics[width=0.8\columnwidth]{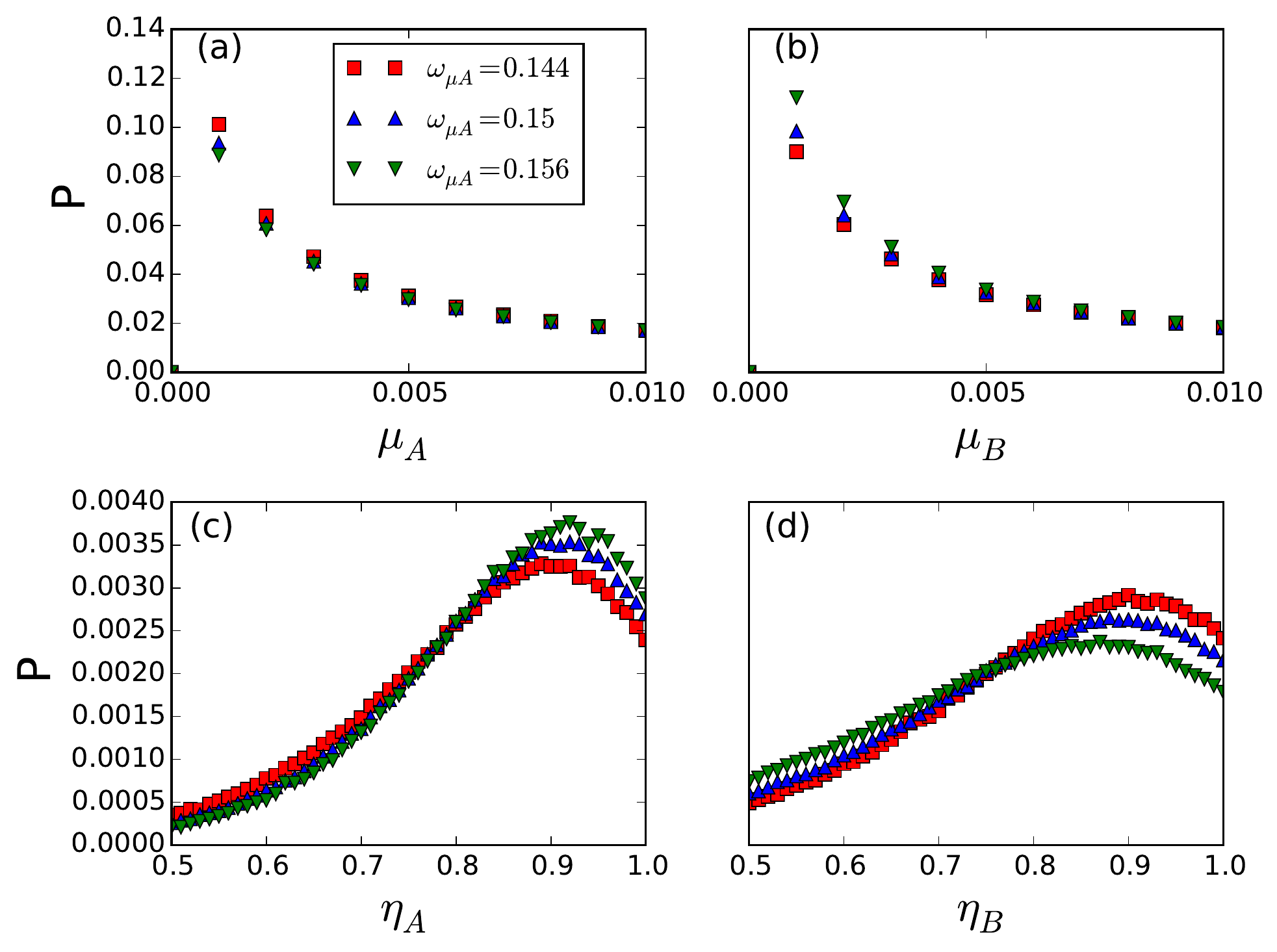} 
\caption{Monte Carlo simulations with direct predator competition:
	The final distribution of predation efficiencies $\eta$ and predator death
	rates $\mu$ after the system has stabilized after $50\,000$ MCS with 
	initial distribution widths $\omega_{\eta A}=0.15$, $\omega_{\eta B}=0.1$, 
	$\omega_{\mu B}=0.1$ and $\omega_{\mu A} \in [0.144, 0.15, 0.156]$; data 
	indicated respectively with red squares, blue triangles up, and green 
	triangles down. 
	(a) and (c) depict the distribution of characters of predator species $A$, 
	while (b) and (d) that of $B$. 
	The interval $[0,1]$ is divided evenly into $1\,000$ histogram bins; the quantity
	$P$ represents the proportion of individuals with rates in the corresponding 
	bins.}
\label{fig13}
\end{figure*}

Biologists rely on direct observation of animals' characters such as beak size 
when studying trait displacement or evolution 
\cite{l1947, bwe1956, g1975, aw1982, gg2006, aald2009, ytprlj2014, jmxl2016}. 
Interspecific competition and natural selection induces noticeable character 
changes within tens of generations so that the animals may alter their phenotype,
and thus look different to their ancestors. 
On isolated islands, native lizards change the habitat use and move to elevated 
perches following invasion by a second lizard kind with larger body size. 
In response, the native subspecies may evolve bigger toepads \cite{sjtp2008}. 
When small lizards cannot compete against the larger ones, character displacement 
aids them to exploit new living habitats by means of developing larger toepads in 
this case, as a result of natural selection.

Interestingly, we arrive at similar observations in our model, where predation 
efficiencies $\eta$ and death rates $\mu$ are allowed to be evolving features of 
the individuals. 
In Fig.~\ref{fig13}, the predation efficiency $\eta$ is initially uniformly set to
$0.5$ for all particles, and the death rate $\mu=0.5$ for all predators (of either
species). 
Subsequently, in the course of the simulations the values of any offspring's 
$\eta$ and $\mu$ are selected from a truncated Gaussian distribution centered at 
their parents' characters with distribution width $\omega_{\eta}$ and 
$\omega_{\mu}$. 
When the system arrives at a final steady state, the values of $\eta$ and $\mu$ 
too reach stationary distributions that are independent of the initial conditions.
We already demonstrated above that smaller widths $\omega$ afford the 
corresponding predator species advantages over the other, as revealed by a larger 
and stable population density.  
In Fig.~\ref{fig13}, we fix $\omega_{\eta A}=0.15$, $\omega_{\eta B}=0.1$,
$\omega_{\mu B}=0.1$, and choose values for 
$\omega_{\mu A} \in [0.144, 0.15, 0.156]$ (represented respectively by red squares,
blue triangles up, and green triangles down), and measure the final distribution of
$\eta$ and $\mu$ when the system reaches stationarity after $50\,000$ MCS. 
Figures~\ref{fig13}(a) and (c) show the resulting distributions for predator 
species $A$, while (b) and (d) those for $B$. 
Since both $\mu$ and $\eta$ are in the range $[0,1]$, we divide this interval 
evenly into $1\,000$ bins, each of length $0.001$. 
The distribution frequency $P$ is defined as the number of individuals whose 
character values fall in each of these bins, divided by the total particle number 
of that species. 
In Fig.~\ref{fig13}(a), the eventual distribution of $\mu_A$ is seen to become 
slightly less optimized as $\omega_{\mu A}$ is increased from $0.144$ to $0.156$
since there is a lower fraction of low $\mu_A$ values in the green curve as 
compared with the red one.  
Since species $A$ has a larger death rate, its final stable population density 
decreases as $\mu_A$ increases. 
In parallel, the distribution of $\eta_A$ becomes optimized as shown in 
Fig.~\ref{fig13}(c), as a result of natural selection: 
Species $A$ has to become more efficient in predation to make up for its 
disadvantages associated with its higher death rates. 
Predator species $B$ is also influenced by the changes in species $A$. 
Since there is reduced competition from $A$ in the sense that its population 
number decreases, the $B$ predators gain access to more resources, thus lending 
its individuals with low predation efficiencies better chances to reproduce, and
consequently rendering the distribution of $\eta_B$ less optimized, see 
Fig.~\ref{fig13}(d). 
This observation can be understood as predator species $B$ needs no longer become
as efficient in predation because they enjoy more abundant food supply. 
In that situation, since species $B$ does not perform as well as before in 
predation, their death rate $\mu_B$ distribution in turn tends to become better 
optimized towards smaller values, as is evident in Fig.~\ref{fig13}(b).  

\subsection{Periodic environmental changes}
\label{subsec4.6}

\begin{figure*}[t]
\centering
\includegraphics[width=0.8\columnwidth]{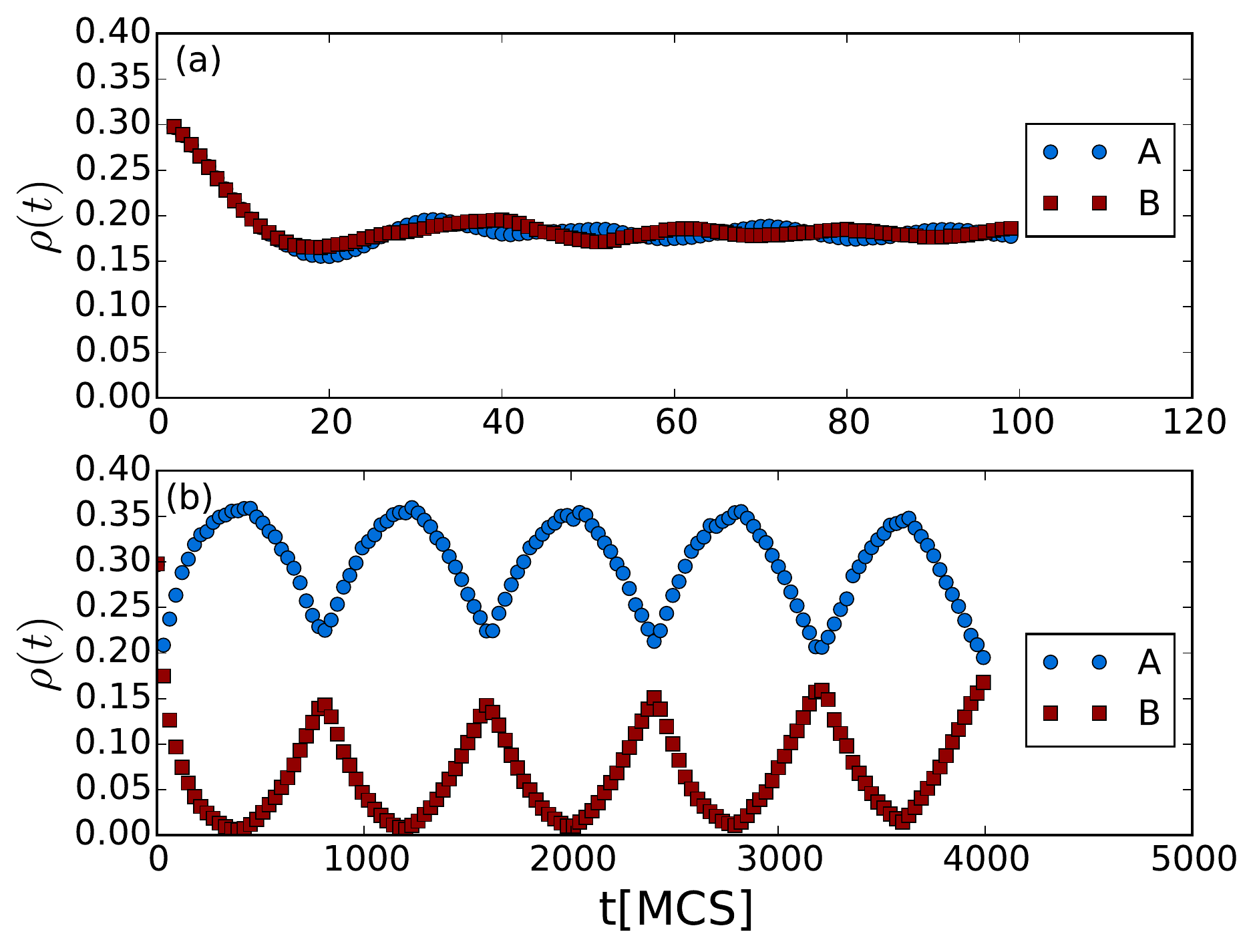} 
\caption{Monte Carlo simulations showing the temporal record of both predator 
	population densities when the distribution widths $\omega_{\eta A}$ and 
	$\omega_{\eta B}$ periodically exchange their values between $0.2$ and $0.3$. 
	The other parameters are set to $\mu_A = \mu_B = 0.125$, $\sigma = 1.0$, and 
	$\omega_C=\omega_{\mu A}=\omega_{\mu B}=0$. 
	The switch periods are $T = 10$ MCS in (a) and $T = 400$ MCS in (b).}
\label{fig14}
\end{figure*}

Environmental factors also play an important role in population abundance. 
There already exist detailed computational studies of the influence of spatial 
variability on the two-species lattice LV model \cite{uu2008, uu2013, uu022013}.
However, rainfall, temperature, and other weather conditions that change in time
greatly determine the amount of food supply. 
A specific environmental condition may favor one species but not others. 
For example, individuals with larger body sizes may usually bear lower temperatures
than small-sized ones. 
Since animals have various characters favoring certain natural conditions, one
may expect environmental changes to be beneficial for advancing biodiversity. 

We here assume a two-predator system with species $A$ stronger than $B$ so that 
the predator $B$ population will gradually decrease as discussed in section 3. 
Yet if the environment changes and turns favorable to species $B$ before it goes 
extinct, it may be protected from extinction. 
According to thirty years of observation of two competing finch species on an 
isolated island ecology \cite{gg2006}, there were several instances when 
environmental changes saved one or both of them when they faced acute danger of 
extinction. 
We take $\omega_{\eta A}$ and $\omega_{\eta B}$ as the sole control parameters 
determining the final states of the system, holding all other rates fixed in our 
model simulations. 
Even though the environmental factors cannot be simulated directly here, we may 
effectively address environment-related population oscillations by changing the
predation efficiency distribution widths $\omega$. 
We initially set $\omega_{\eta A}=0.2$ and $\omega_{\eta B}=0.3$, with the other 
parameters held constant at $\mu_A = \mu_B = 0.125$, $\sigma = 1.0$, and 
$\omega_C=\omega_{\mu A}=\omega_{\mu B}=0$. 
In real situations the environment may alternate stochastically; in our idealized
scenario, we just exchange the values of $\omega_{\eta A}$ and $\omega_{\eta B}$ 
periodically for the purpose of simplicity. 
The population average of the spontaneous death rate is around $0.02$, therefore 
its inverse $\approx 50$ MCS yields a rough approximation for the individuals'
typical dwell time on the lattice. 
When the time period $T$ for the periodic switches is chosen as $10$ MCS, which is
shorter than one generation's life time, the population densities remain very close
to their identical mean values, with small oscillations; see Fig.~\ref{fig14}(a). 
Naturally, neither species faces the danger of extinction when the environmental 
change frequency is high. 
In Fig.~\ref{fig14}(b), we study the case of a long switching time $T = 400$ MCS,
or about eight generations. 
As one would expect, the $B$ population abundance decreases quickly within the 
first period. 
Before the $B$ predators reach total extinction, the environment changes to in turn
rescue this species $B$. 
This example shows that when the environment stays unaltered for a very long time, 
the weaker species that cannot effectively adapt to this environment would 
eventually vanish while only the stronger species would survive and strive. 
When the time period $T$ close matches the characteristic decay time $t_c$, see 
Fig.~\ref{fig14}(b), one observes a resonant amplification effect with large 
periodic population oscillations enforced by the external driving.

\section{Summary}
\label{sec5}

In this paper, we have used detailed Monte Carlo simulations to study an ecological
system with two predator and one prey species on a two-dimensional square lattice. 
The two predator species may be viewed as related families, in that they pursue 
the same prey and are subject to similar reactions, which comprise predation, 
spontaneous death, and (asexual) reproduction. 
The most important feature in this model is that there exists only one mobile and
reproducing food resource for all predators to compete for. 
We have designed different model variants with the goal of finding the key 
properties that could stabilize a three-species coexistence state, and thus 
facilitate biodiversity in this simple idealized system. 
We find no means to obtain such coexistence when all reaction rates are fixed or 
individuals from the same species are all homogeneous, which clearly indicates the 
importance of demographic variability and evolutionary population adaptation. 
When dynamical optimization of the individuals in the reproduction process is 
introduced, they may develop various characters related to their predation and 
reproduction efficiencies. 
However, this evolutionary dynamics itself cannot stabilize coexistence for all 
three species, owing to the fixed constraint that both predator kinds compete for 
the same food resource.  
In our model, direct competition between predator species is required to render a 
three-species coexistence state accessible, demonstrating the crucial importance of 
combined mutation, competition, and natural selection in stabilizing biodiversity.

We observe critical slowing-down of the population density decay near the predator 
extinction thresholds, which also serves as an indicator to locate the coexistence 
region in parameter space. 
When the system attains its quasi-steady coexistence state, the spatial properties 
of the particle distribution remain stable even as the system evolves further. 
Character displacements hence occur as a result of inter-species competition and 
natural selection in accord with biological observations and experiments.
Through comparison of the coexistence regions of the full lattice model and its
zero-dimensional representation, we find that spatial extent may in fact reduce 
the ecosystem's stability, because the two predator species can effectively block 
each other from reaching their prey. 
We also study the influence of environmental changes by periodically switching the 
rate parameters of the two competing predator species. 
The system may then maintain three-species coexistence if the period of the 
environmental changes is smaller than the relaxation time of the population density 
decay.
Matching the switching period to the characteristic decay time can induce 
resonantly amplified population oscillations.

Stable coexistence states with all three species surviving with corresponding 
constant densities are thus only achieved through introducing both direct predator 
competition as well as evolutionary adaptation in our system. 
In sections 3 and 4, we have explored character displacement without direct 
competition as well as competition without character displacement, yet a stable 
three-species coexistence state could not be observed in either case. 
Therefore it is necessary to include both direct competition and character 
displacement to render stable coexistence states possible in our model. 
However, both predator species $A$ and $B$ can only coexist in a small parameter
interval for their predation efficiency distribution widths $\omega$, because they 
represent quite similar species that compete for the same resources. 
In natural ecosystems, of course other factors such as distinct food resources 
might also help to achieve stable multi-species coexistence.  

\section*{Acknowledgments}
The authors are indebted to Yang Cao, Silke Hauf, Michel Pleimling, Per Rikvold
and Royce Zia for insightful discussions. U.D. was supported by a Herchel Smith
Postdoctoral Fellowship.

\section*{References}

%\bibliography{mybibfile}

\end{document}